\documentclass[12pt, titlepage]{article}
\usepackage[margin=1in]{geometry} 
\usepackage{setspace}
\usepackage{enumerate}
\usepackage{natbib}
\usepackage{graphicx}
\usepackage{amssymb}
\usepackage{epstopdf}
\usepackage{amsmath}
\usepackage{url}
\usepackage[labelfont=bf]{caption}
\usepackage{pdflscape}
\usepackage{afterpage}
\usepackage{longtable}
\usepackage{hyperref} 
\hypersetup{
    colorlinks,
    citecolor=black,
    filecolor=black,
    linkcolor=blue,
    urlcolor=black
}
\hypersetup{linktocpage}

\title{Probabilistic Projection of Subnational Total Fertility Rates}
\author{Hana \v{S}ev\v{c}\'{\i}kov\'{a}\footnotemark[1], Adrian E. Raftery\footnotemark[1], Patrick Gerland\footnotemark[2]\\[20pt]
\footnotemark[1]{University of Washington, Seattle} \\
\footnotemark[2]{United Nations Population Division, New York}
}
\date{\today}

\begin{document}
\pagenumbering{roman}

\maketitle

\begin{abstract}
We consider the problem of probabilistic projection of the total fertility
rate (TFR) for subnational regions. We seek a method that is consistent
with the UN's recently adopted Bayesian method for probabilistic TFR 
projections for all countries, and works well for all countries.
We assess various possible methods using subnational TFR data for 47 countries.
We find that the method that performs best in terms of out-of-sample 
predictive performance and also in terms of reproducing the within-country
correlation in TFR is a method that scales the national trajectory by 
a region-specific scale factor that is allowed to vary slowly over time.
This supports the hypothesis of Watkins (1990, 1991) that within-country
TFR converges over time in response to country-specific factors,
and extends the Watkins hypothesis to the last 50 years and to a much
wider range of countries around the world.\\[3mm]
\noindent {\it Keywords:} {Total fertility rate, Subnational projections, Autoregressive model, Bayesian hierarchical model, Scaling model, Correlation.}
\end{abstract}

\pagenumbering{arabic}

\tableofcontents

\clearpage

\baselineskip=18pt

\section{Introduction}
The United Nations Population Division issued official probabilistic
population projections for all countries for the first time in 2015 
\citep{UN2015ppp}, using the methodology described by \citet{RafteryLi&2012}.
One of the key components of the projection methodology is a Bayesian 
hierarchical model for the total fertility rate (TFR) in all countries 
\citep{Alkema&2011,RafteryAlkemaGerland2013,FosdickRaftery2014}.

Population projections for subnational administrative units, such
as provinces, states, counties, regions or d\'epartements 
(hereafter all referred to simply as regions), are of great interest to
national and local governments for planning, policy and decision-making
\citep{Rayer&2009}.
A common current practice is to generate subnational projections 
deterministically by scaling national projections \citep{USCensusBureau}.
Specifically, the US Census Bureau provides a workbook for users to generate 
subnational TFR projections for up to 32 regions. The method requires the
user to enter an ultimate TFR level (lower asymptote) to which the regional 
TFR converges, and a deterministic projection of the national TFR. 
The subnational TFR is then projected in such a way that it approaches 
the target TFR with the same rate as the national TFR approaches this target.
This method does not yield probabilistic projections. 

In this paper we try to address one aspect of the problem,
namely probabilistic subnational projections of TFR.
Methods for probabilistic subnational projections have been 
developed for individual countries or parts of countries
\citep{SmithSincich1988,Tayman&1998,ReesTurton1998,GullicksonMoen2001,Gullickson2001,Lee&2003,SmithTayman2004,WilsonBell2007,Rayer&2009,Raymer&2012,Wilson2013}; 
for a review see \citet{Tayman2011}.
Our ultimate goal is to extend the UN method for probabilistic
projections for all countries to a method for subnational probabilistic
projections that is consistent across countries and 
works well for all regions of all countries.

We contrast two broad approaches to subnational probabilistic projection of TFR.
One approach is a direct extension of the UN method \citep{Alkema&2011}
to subnational data, effectively treating the country in the same way
the UN model treats the world, and treating the regions in the
same way the UN model treats the countries. \citet{Borges2015} proposed
an approach along these lines for the provinces of Brazil. 

The other approach 
is motivated by the observation of 
Watkins (1990, 1991) \nocite{Watkins1990,Watkins1991} 
that within-country variation in TFR in Europe decreased over the 
period of the fertility transition, between 1870 and 1960.
This observation has been confirmed for a more recent period for 
the German-speaking countries \citep{Basten2012}, to some extent for India 
\citep{ArokiasmyGoli2012,Wilson&2012}, while the evidence is more equivocal for 
the United States \citep{OConnell1981}.
Watkins posited that this was due to increased integration of national markets, 
expansion of the role of the state, and nation-building in the form
of linguistic standardization  over this period.
\citet{Calhoun1993} argued that, of these three mechanisms, only
linguistic standardization clearly supported her argument.
However, some support for the 
importance of the role of the nation state for fertility
is provided by the fact that nation states have specific and different 
policies aimed at affecting fertility rates \citep{Tomlinson1985,Chamie1994}, 
and some of these policies have been shown to be effective 
\citep{Kalwij2010,Luci-Gruelich2013}.
Note that \citet{Klusener2013} investigated subnational convergence
of {\it non-marital} fertility in Europe in recent decades, and found that 
within-country variation increased, in contrast with the trends noted
by other authors for overall fertility. Here we consider only overall fertility.

One question is then whether the direct extension of the UN method for 
countries to the subnational context adequately
accounts for this tendency of TFR to converge within countries over time.
Note that this extension of the UN method does predict within-country 
convergence of fertility rates over time during the fertility transition; 
the question is whether it adequately accounts for this convergence.

To investigate this question, we consider a different general approach,
which starts from the national probabilistic projections produced by the
UN method, and then scales them for each region by a scaling
factor that varies stochastically, but stays relatively constant. 
This induces more within-country correlation than the direct extension
of the UN method. It could be viewed as a probabilistic extension of the
method currently used by the U.S.~Census Bureau.
It is also related to the method of \citet{Wilson2013}, but with
some significant differences.

We apply these methods to subnational data on total fertility for 
47 countries over the period 1950--2010.
We compare our two approaches and several variants in terms of out-of-sample
predictive performance. The results shed some light on the Watkins
hypothesis of increasing within-country correlation, as well providing
some guidance on how to carry out subnational probabilistic TFR projection.

Note that there is a substantial literature on convergence of fertility
rates in different countries to one another, with different conclusions 
argued for
\citep{Wilson2001,Wilson2004,Reher2004,Reher2007,Dorius2008,Wilson2011}. 
Our work here has implications for within-country fertility 
convergence, but is agnostic about fertility convergence between countries,
and so does not have implications for global fertility convergence,
for example.

The paper is organized as follows. We first describe the data used in this study and review the model for national probabilistic projections.  
We then introduce our proposed methodology for subnational probabilistic
projections, and present the results. The paper concludes with a discussion.

\section{Data}
\label{sec:data}
\begin{figure}
\includegraphics[width=\textwidth]{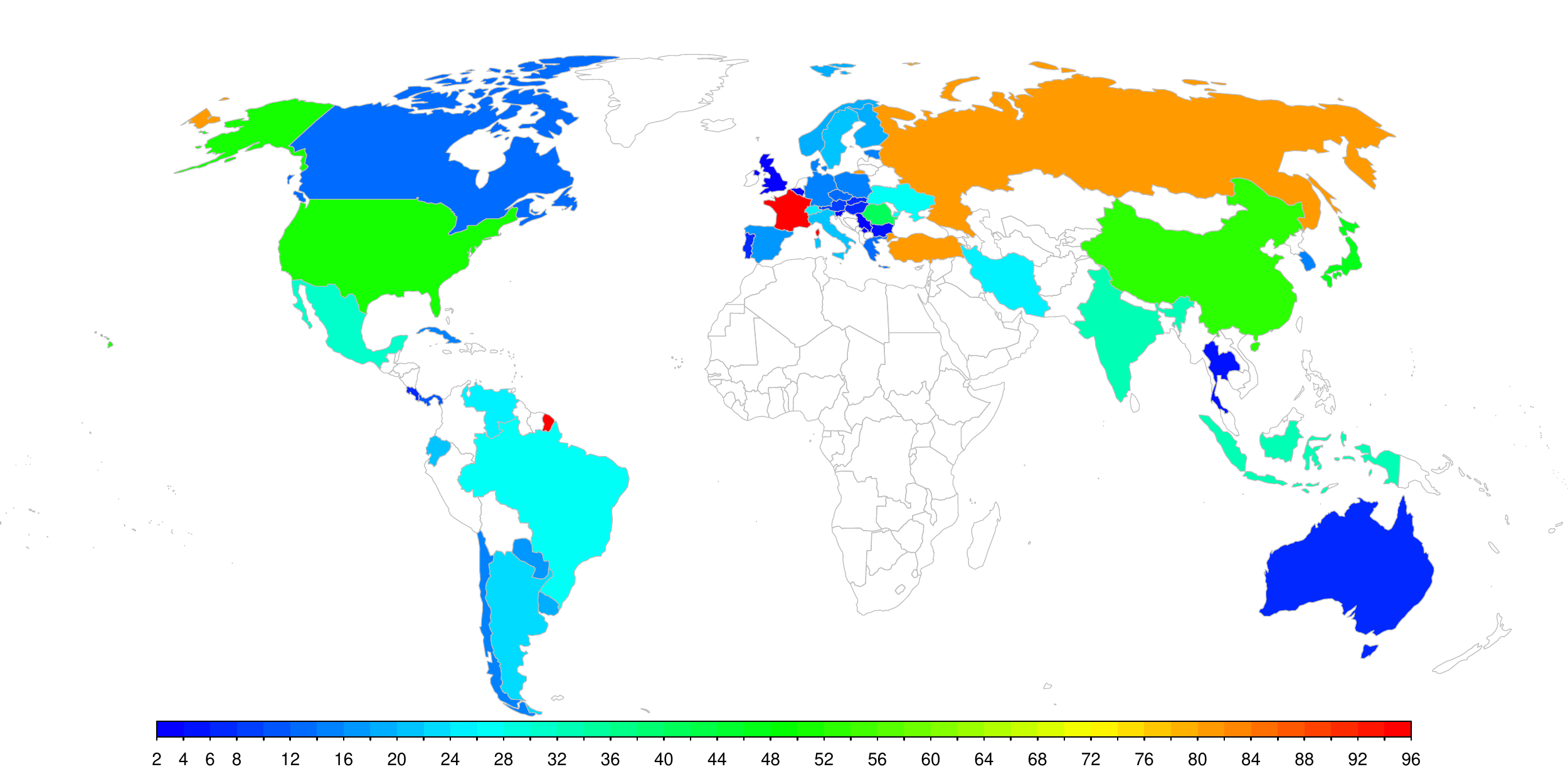}
\caption{\label{fig:data-map}Map of 47 countries with subnational TFR data. The color scale shows the number of regions for each country, which ranges 
from 2 to 96.}
\end{figure}

We use subnational data on the TFR for 47 countries (13 in the Americas,
nine in the Asia-Pacific region, and 25 in Europe), corresponding to 1,092
regions for the period 1950--2010, collected by the United Nations 
Population Division. Each country analyzed had a population over one 
million and a national average TFR below 2.5 in 2010--2015. 
The geographical level selected for each country was the one with 
available data for the longest comparable time series.
The dataset covers 4.9 billion people.
Fig.~\ref {fig:data-map} shows the numbers of regions 
for each country, which range from two for Slovenia to 96 for France. 
The data include countries from all the inhabited continents except Africa. 
The data sources are shown in Appendix Table \ref{tab:datasource}.

Fig.~\ref{fig:data} shows an example of the data for four countries (USA, India, Brazil and Sweden). It illustrates that the data vary with respect to the correlation between regions. It also shows that the data started later
than 1950 for some regions.
In the figure, the national TFR from \citet{UN2013} is shown as a black curve.

\begin{figure}
\begin{minipage}{0.49\textwidth}
\includegraphics[scale=0.33]{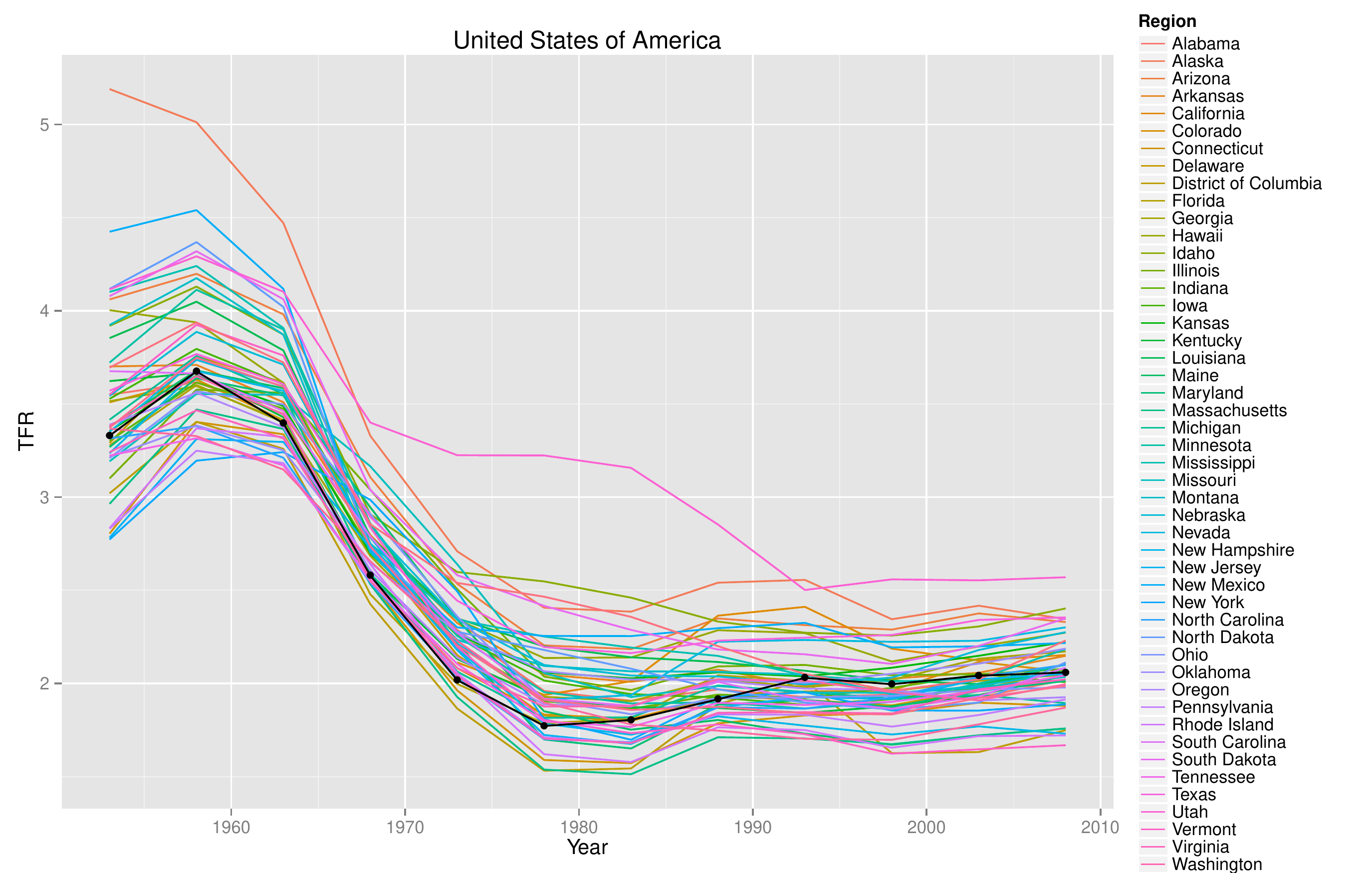}
\end{minipage}
\begin{minipage}{0.49\textwidth}
\includegraphics[scale=0.33]{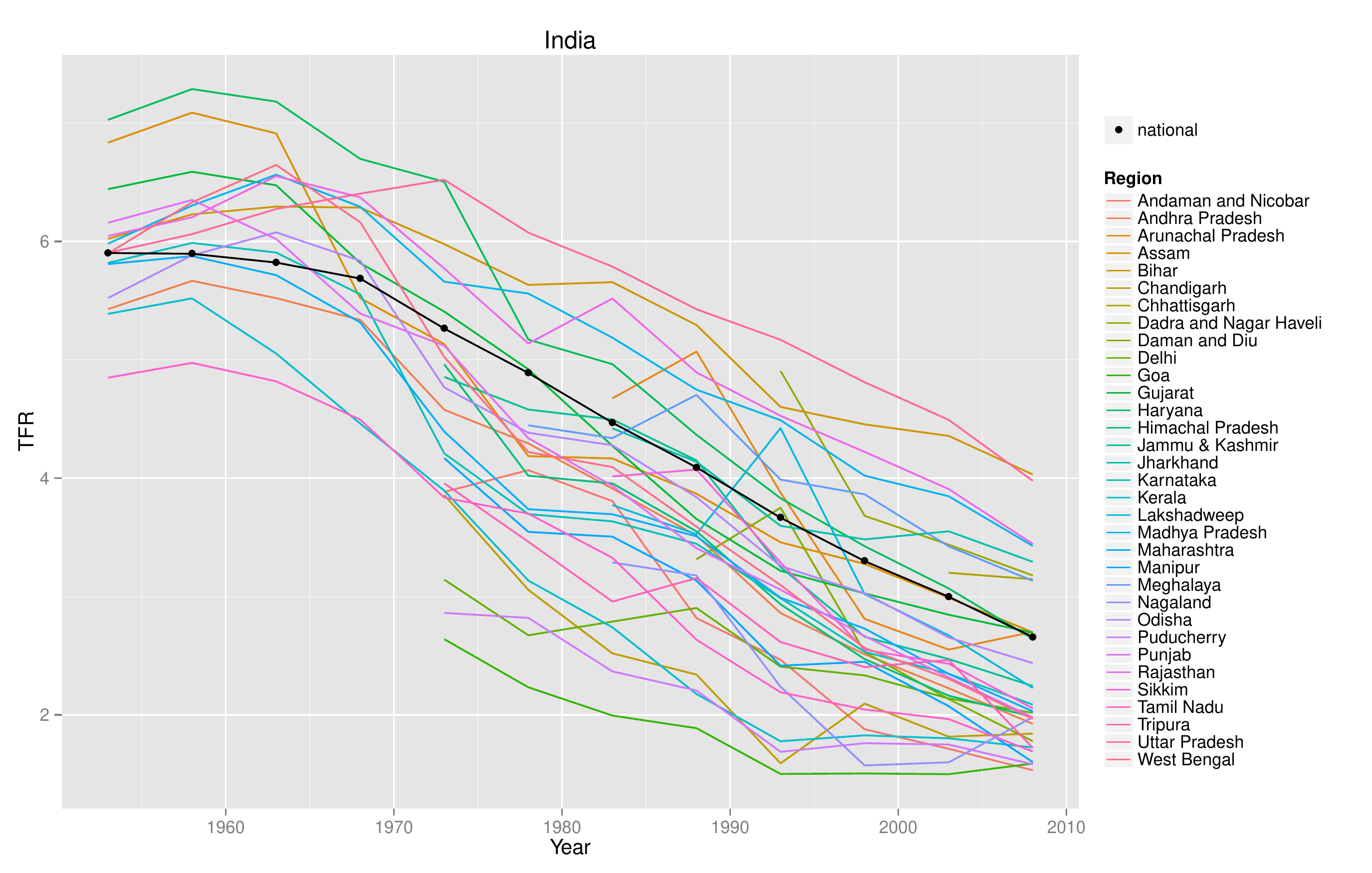}
\end{minipage}

\begin{minipage}{0.49\textwidth}
\includegraphics[scale=0.33]{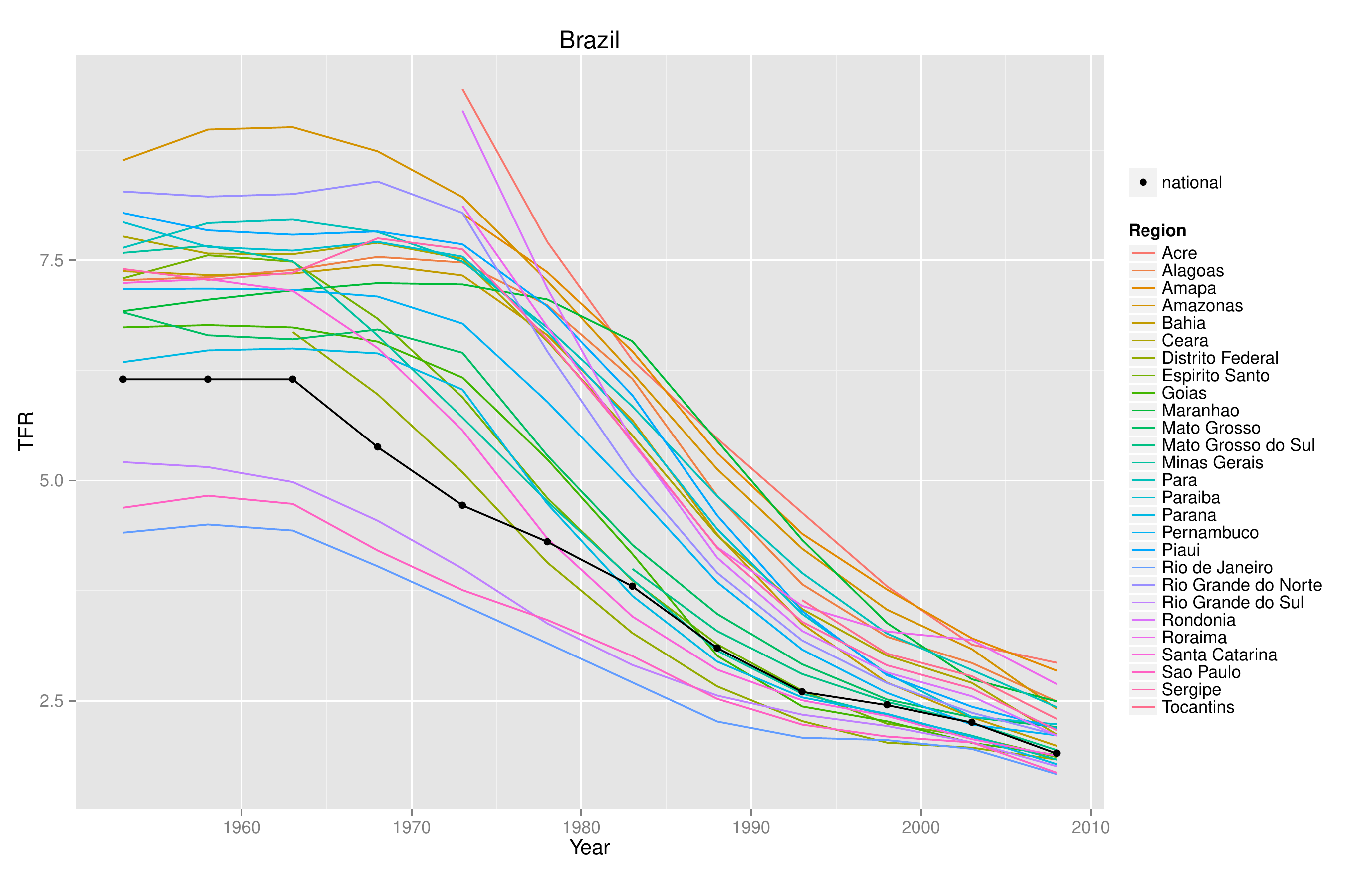}
\end{minipage}
\begin{minipage}{0.49\textwidth}
\includegraphics[scale=0.33]{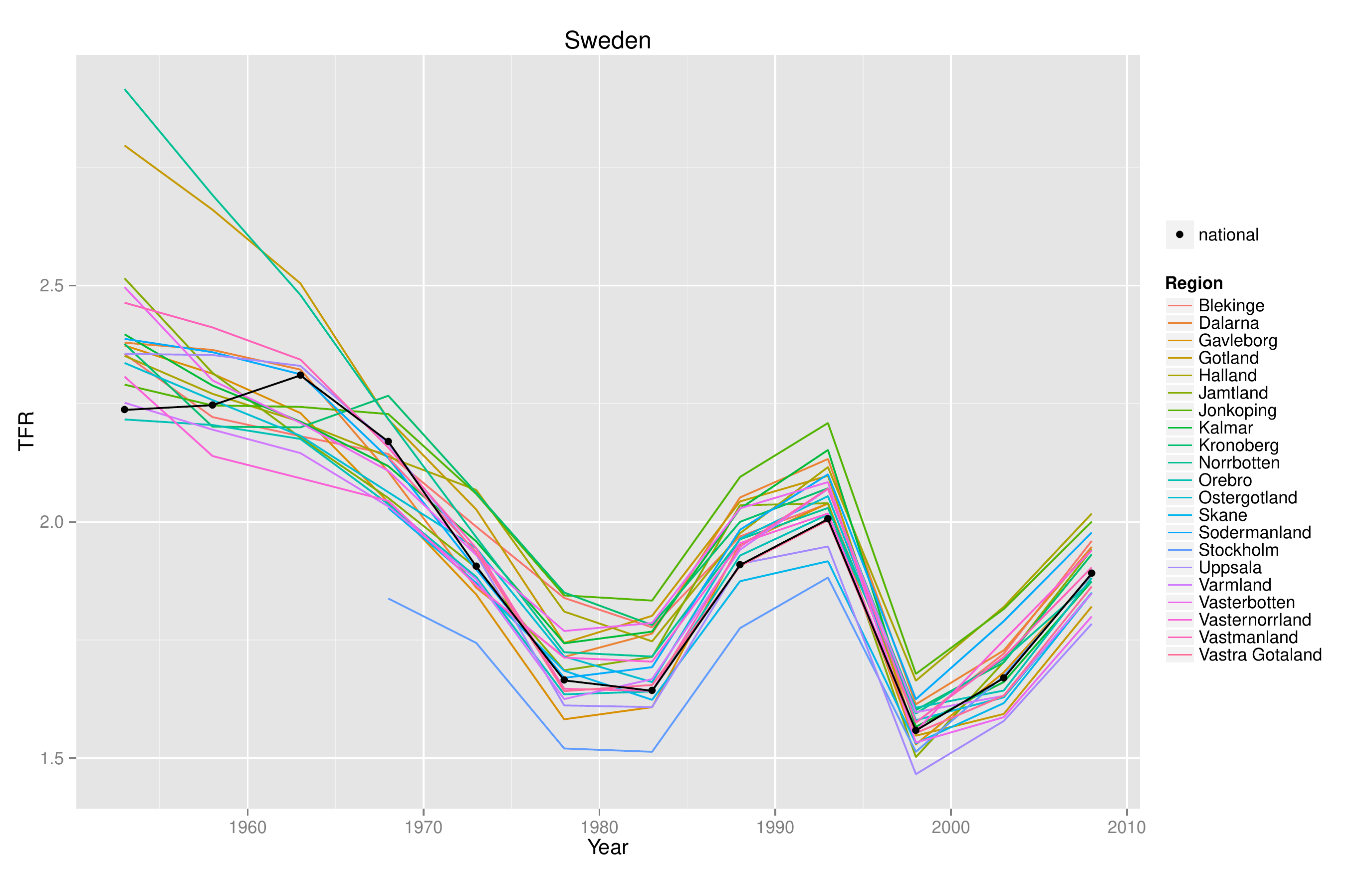}
\end{minipage}
\caption{\label{fig:data} Observed data for regions of four countries, namely the USA, India, Brazil, and Sweden. The national TFR is shown by the black curve.}
\end{figure}

\section{Review of the national Bayesian hierarchical model}
\label{sec:national}
Our starting point for developing a methodology for subnational projections is the probabilistic model for projecting national TFR proposed by  \citet{Alkema&2011}, which has now been adopted by the UN for its official projections.
We start by summarizing the main ideas of this Bayesian hierarchical model (BHM). More detail can be found in \citet{Alkema&2011} and \citet{RafteryAlkemaGerland2013}.

\begin{figure}
\begin{minipage}{0.55\textwidth}
\includegraphics[width=\textwidth]{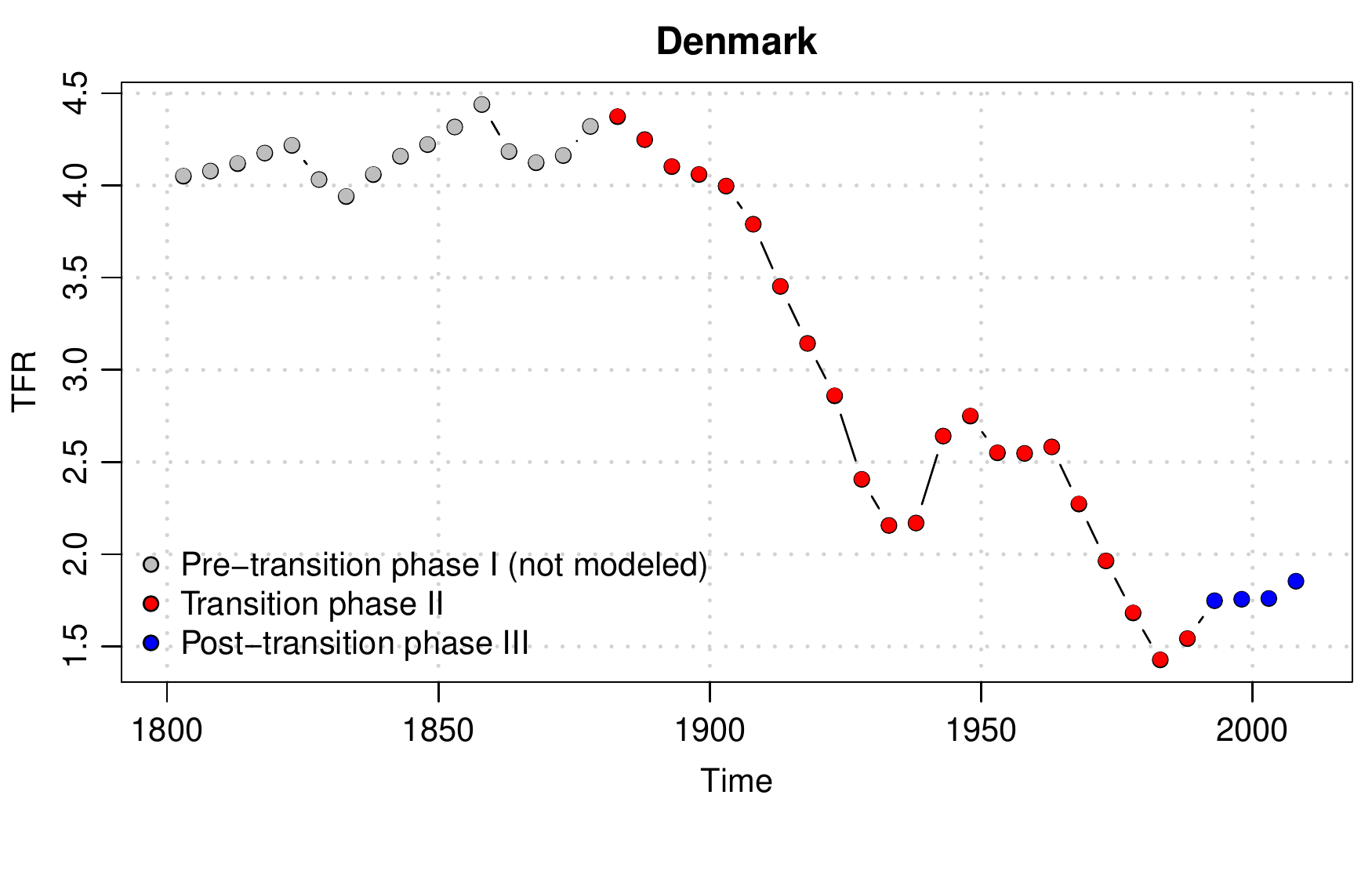}
\end{minipage}
\begin{minipage}{0.43\textwidth}
\includegraphics[width=\textwidth]{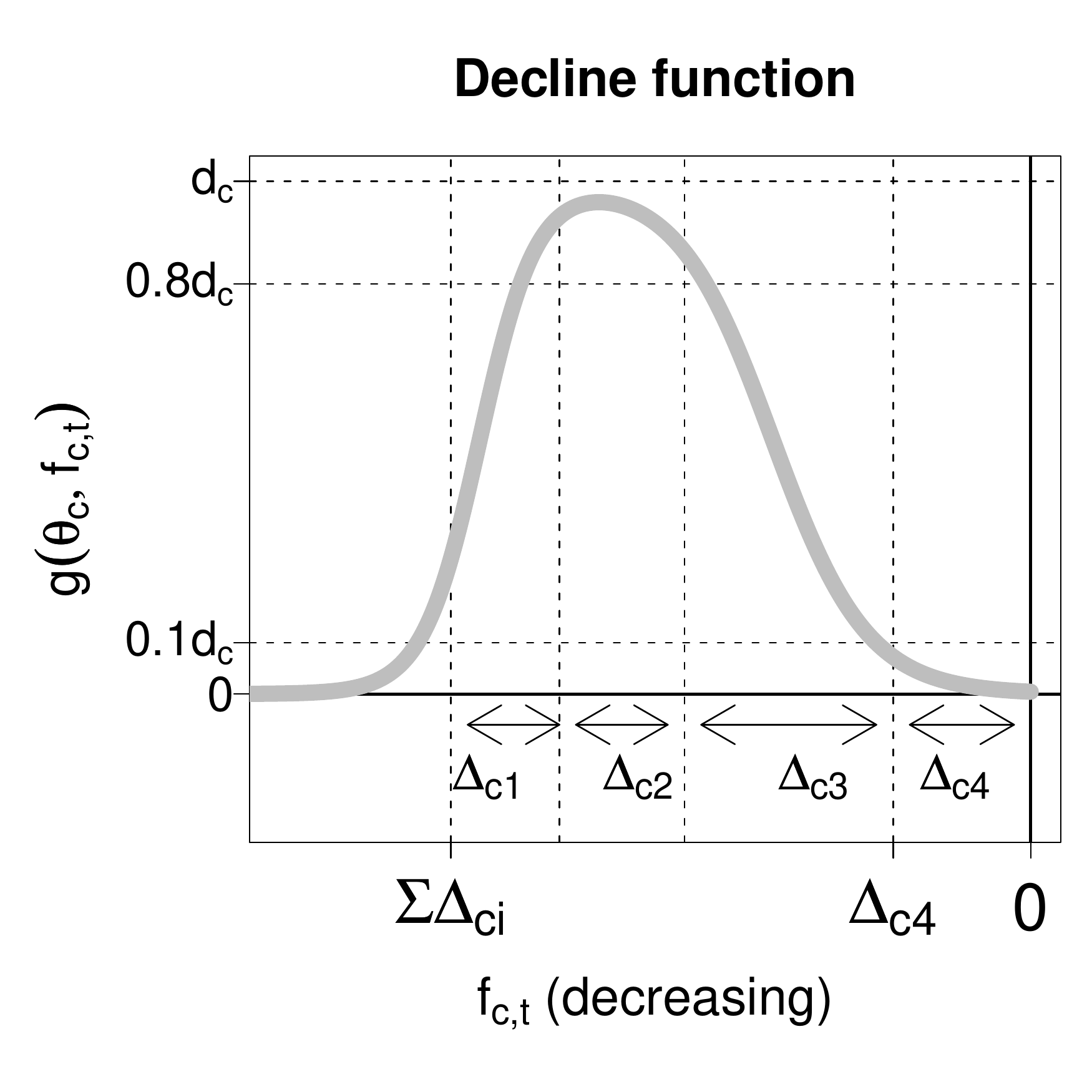}
\end{minipage}
\caption{\label{fig:tfr-phases-dl} Left panel: Three phases of the typical TFR evolution on the example of Denmark. Right panel: Cartoon of a double logistic decline curve for country $c$ with its parameters defining the shape. $f_{c,t}$ on the $x$ axis denotes the TFR; $g(\theta_c,f_{c,t})$ on the $y$ axis denotes the first order difference in TFR.}
\end{figure}

The model is based on standard fertility transition theory
(e.g. Hirschman 1994) \nocite{Hirschman1994}, and is compatible with 
almost all versions of this in the literature.
It distinguishes three phases in the evolution of a country's 
fertility over time, depicted in the left panel 
of Fig.~\ref{fig:tfr-phases-dl} for the example of Denmark.
Phase I (grey dots) precedes the beginning of the fertility transition and is characterized by high fertility that is stable or increasing. This phase is not modeled as all or nearly all countries have completed this phase. During Phase II, or the transition phase (red dots in the figure), fertility declines from high levels to below the replacement level of 2.1 children per woman. Phase III is the post-fertility transition period (blue dots), during which fertility
fluctuates at low levels, possibly recovering towards the replacement level.

To model the fertility declines in each five-year period during Phase II, a double logistic decline function is used. An example of this function is shown in the right panel of Fig.~\ref{fig:tfr-phases-dl}. The function is parametrized by a set of country-specific parameters that define the shape of the country's decline curve. Those parameters are drawn from a world distribution. The resulting BHM is estimated using Markov chain Monte Carlo (MCMC).

Phase III is modeled using a Bayesian hierarchichal first-order  
autoregressive, or AR(1),  process of the form:
\[
f_{c,t+1}-\mu_c =  \rho_c (f_{c, t}-\mu_c) + \varepsilon_{c,t} , \quad \text{with} \quad \varepsilon_{c,t} \stackrel{iid}{\sim} N(0, \sigma_{\varepsilon}^2) .
\]
It implies that fertility for country $c$ has a country-specific long-term mean, $\mu_c$, and autoregressive parameter, $\rho_c$, which are assumed to be
drawn from a world distribution. The parameters of this world distribution
in turn have a joint prior distribution, thus defining a three-level
hierarchical model, where the three levels are the observation, the 
country and the world. The resulting model is again estimated by MCMC.

The process of estimating Phase II and Phase III parameters results in a set of country-specific decline curves and a set of country-specific AR(1) parameter pairs. Unlike decline curves which can be estimated for all countries, not all countries have experienced Phase III, in which cases the country-specific long-term means and autoregressive parameters cannot be estimated. In such cases, the ``world" means and autoregressive parameters  are used. The estimated parameters are then used to generate a set of future TFR trajectories yielding probabilistic TFR projections for all countries of the world.

\section{Methods for subnational projections}
\label{sec:methods}
Ideally, we seek a method for generating probabilistic subnational TFR projections that reflects the literature and theory of fertility transitions, is based on the national methodology used by the UN and described above, works well for all countries, is as simple if possible, and yields correlations between regions that are similar to the correlations in the observed data. 

We first describe a simple Scale method that provides an initial probabilistic
extension of methods used by the U.S.~Census Bureau and other national agencies.
This simple approach works well from many points of view, but it 
does not allow for the possibility of crossovers between regions,
whereas in fact these do happen. We therefore elaborate this model
to allow the scale factor to change stochastically, but slowly over time,
yielding the so-called Scale-AR(1) method.
Finally we describe a quite different approach, called the one-directional BHM,
which directly generalizes the national approach to the subnational context, 
allowing regions to vary more freely within a country.

\subsection{Scale Method}
\label{sec:scale}
We start with a simple intuitive scale method where, for each trajectory
from the probabilistic projection,  the regional TFR is simply a product 
of the simulated national TFR and a time-independent but region-specific 
scale factor.

Let $f_{c,t,i}$ denote the national TFR projection for country $c$ at time $t$ from trajectory $i$, simulated from its posterior distribution as described
above. We model $f_{r_c,t,i}$,  the TFR for region $r_c$ of country $c$ at time $t$ in the $i$-th trajectory, by
\begin{equation}
\label{eq:TFRscalesimple}
f_{r_c,t,i} = \alpha_{r_c} f_{c,t,i} ,
\end{equation}
where $\alpha_{r_c}$ denotes the regional scaling factor derived from the last observed (present) time period denoted by $P$:
\begin{equation}
\label{eq:TFRscalesimple-alpha}
\alpha_{r_c} = f_{r_c,t=P} / f_{c,t=P} .
\end{equation}
Note that $\alpha_{r_c}$ is the same for all trajectories.
This method yields a set of regional trajectories $f_{r_c,t,i}$ 
and thus yields probabilistic projections of the regional TFRs, $f_{r_c,t}$.

Our numerical experiments, described below, indicated that this simple method
performed surprisingly well. However, it also has a serious drawback.
Scaling by a constant factor yields a perfect correlation, 
i.e. it does not allow for the possibility of crossovers between regions 
over time. However, such crossovers do happen, and the scale method says
that they are impossible, which is not fully satisfactory.

\subsection{Scale-AR(1)}
\label{sec:ar1scale}
To avoid this drawback, and modify the scale method so as to allow for
the possibility of crossovers,
we propose a variation of the simple Scale method where we model the regional scale factor using a first-order  autoregressive, or AR(1),  process:

\begin{equation}
\label{eq:AR1}
\alpha_{r_c,t} - 1 =  \phi(\alpha_{r_c,t-1}-1)  + \varepsilon_{r_c,t}, \quad \text{with} \;\; \varepsilon_{r_c,t} \stackrel{iid}{\sim}  N(0, \sigma_{c}^2) .
\end{equation}
The regional TFR $f_{r_c,t,i}$ is then derived as in (\ref{eq:TFRscalesimple}) with the additional lower bound restriction, $f_{r_c,t,i} > 0.5$.

This model implies that the scaling factor will fluctuate around one in the long term. Regardless of its initial value, it will converge to a distribution
that is centered around one,  and the rate of convergence is determined by the $\phi$ parameter. 
We use the following settings for the model parameters, estimated from the
data for all 47 countries available:
\begin{eqnarray}
\label{eq:AR1est}
\phi &=& 0.925 , \\
\sigma^2_{c}&=& \min\{ \sigma^2, (1-\phi^2)\text{Var}_{r \in R_c}(\alpha_{r,t=P})\} , \label{eq:AR1est-sigmac}\\
\sigma &=& 0.0452 , \label{eq:AR1est-sigma}
\end{eqnarray}
where $P$ again denotes the present time period and $R_c$ denotes the set of regions of country $c$. The minimum restriction in (\ref{eq:AR1est-sigmac}) ensures that the variation of $\alpha_{\cdot, t}$ across regions is not larger than the variation in the last observed time period, in line with the Watkins hypothesis
and the long-term observed data.
Details of how these parameters were estimated are given in the appendix.

This method is related to the method proposed by \citet{Wilson2013},
but there are some significant differences that are discussed in the
Discussion section.


\subsection{One-directional BHM}
\label{sec:one-d-BHM}
\begin{figure}
\begin{center}
\includegraphics[scale=0.5]{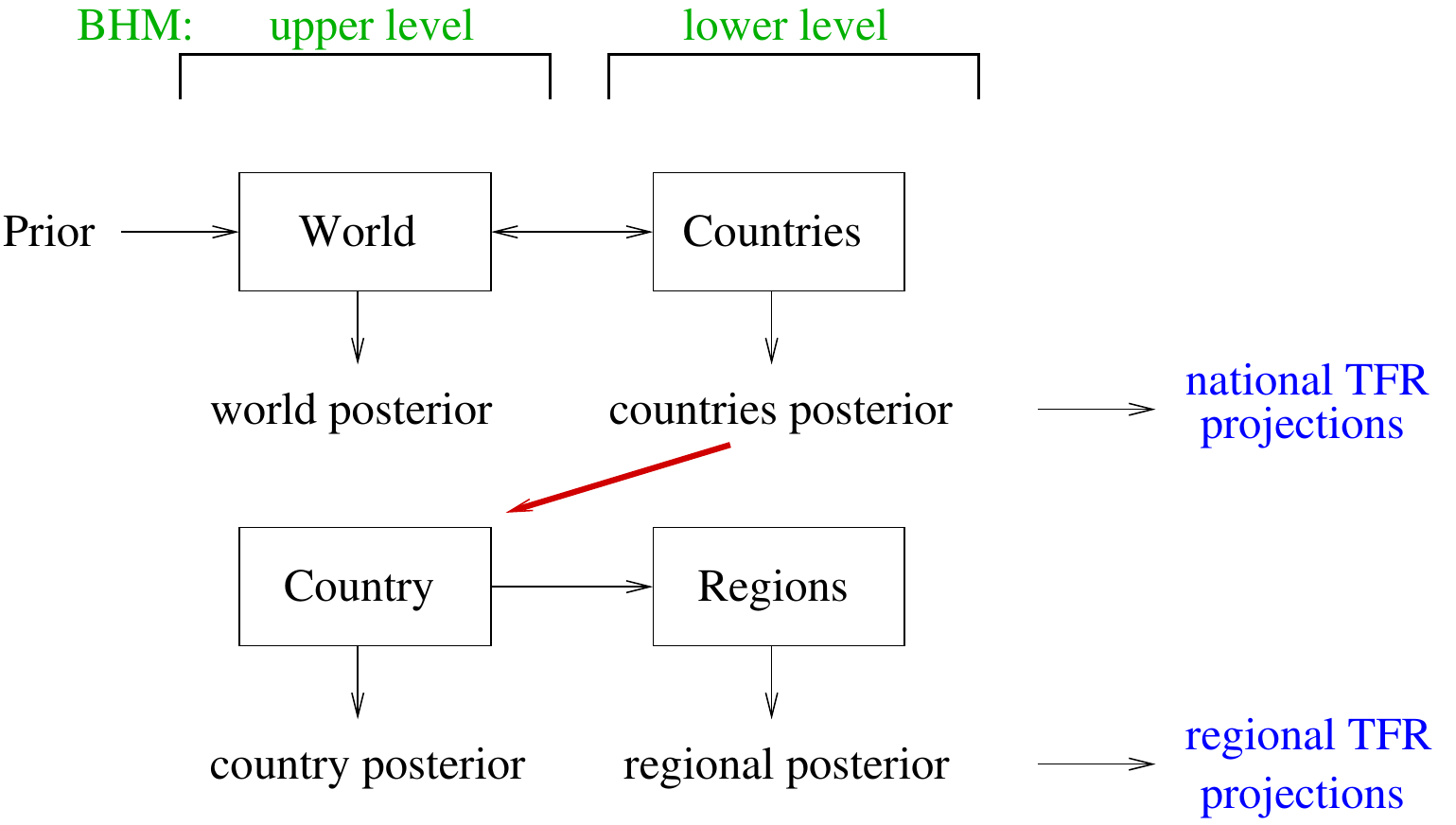}
\end{center}
\caption{\label{fig:BHM1d} One-directional BHM for the  subnational model.}
\end{figure}

Next, we consider an extension of the world three-level BHM which is depicted in Fig.~\ref{fig:BHM1d}. The three levels of this model are the world level,
country level and time point or observation level. 
(The time point level is not shown in the figure). 
In the world version, information from all countries is combined into the
world level, which in turn influences the country level, yielding a 
two-directional BHM.
The prior distribution of the hyperparameters is vague for most parameters, 
but also reflects expert knowledge in some cases. 
The model yields a posterior distribution of the 
world parameters, and the country-specific parameters, 
which is then used to generate the national projections. 

Our extension has a similar setup, but moves down by one level of geography and works in one direction only. 
Thus the top level of our national model is the country,
the next level is the region, and the bottom level is the time point.
The upper level of our model corresponds to the country level of the world model, that is, we carry over the country-specific posterior from a world simulation and use it as the distribution of the hyperparameters in our national model (red arrow in Fig.~\ref{fig:BHM1d}). On the lower level, data from all regions of a country are handled individually. The estimation of the regional parameters is informed by the hyperparameters, but the regional level does not influence the country level of the model.
The resulting regional posterior distribution is used to project subnational TFR. 

Note that
many countries do not have historical data on Phase III because they have not yet reached this stage, and so in these cases the country posterior is the same as the world posterior. As a result, all regions of those countries inherit the ``world" Phase III parameters.

\subsubsection{Correlation between regions}
\label{sec:correlation}
For aggregating TFR over sets of regions, for example for deriving country's averages, it is important to capture correlation in model errors between regions, as was done by \citet{FosdickRaftery2014} for capturing correlation between countries.

We will model the forecast errors as follows:
\begin{equation}
\boldsymbol {\varepsilon_t} \sim N(0, \Sigma_t = \boldsymbol {\sigma_t'} \boldsymbol{A} \boldsymbol {\sigma_t}) ,
\label{eq-epsilon}
\end{equation}
where $\boldsymbol {\sigma_t}$ is a vector consisting of the forecast standard deviations for each region. For Phase II this is the standard deviation of the errors in the double logistic model and for Phase III it is the standard deviation of the AR(1) model. In (\ref{eq-epsilon}), $\boldsymbol{A}$ is a matrix where each element $A_{r,s}$ corresponds to the correlation between the model errors of region $r$ and $s$ over all time periods. 

Let $f_{r,t}$ denote the observed TFR for region $r$ at time $t$. 
We denote by $e_{r,t}$ the normalized forecast error, namely 
the forecast error divided by its standard deviation.
The normalized forecast error $e_{r,t}$ is estimated as follows:
\begin{itemize}
\item {\bf Phase II}: For each value $g_{r,t,i}$ of a double logistic (DL) trajectory $i$ and the standard deviation of DL $\sigma_{r,i}$ take $d_{r,t,i} = (f_{r,t} - g_{r,t,i})/\sigma_{ri}$. Then $e_{r,t}$ is the mean of $d_{r,t,i}$ over $i$.
\item {\bf Phase III}: For each value $h_{r,t,i}$ of a phase III trajectory $i$ and the standard deviation of these trajectories $\sigma_{\varepsilon,r,i}$, take the difference $d_{r,t,i} = (f_{r,t} - h_{r,t,i})/\sigma_{\varepsilon,r,i}$. Then, $e_{r,t}$ is the mean of $d_{r,t,i}$ over $i$.
\end{itemize}
We define the correlation matrix $\boldsymbol{A}$ as 
\begin{equation}
\boldsymbol{A} = \frac{\bar{T}-1}{\bar{T}} \boldsymbol{\tilde{A}} + \frac{1}{2\bar{T}}
\end{equation}
where $\boldsymbol{\tilde{A}}$ is a truncated correlation matrix made positive definite, and $\bar{T}$ is the average number of time periods per region. 
Here $\boldsymbol{A}$ has an approximate Bayesian interpretation as an 
approximation of the posterior mean with a uniform $U[0,1]$ on the 
correlations. Note that $\boldsymbol{A}$ is positive definite.
The appendix contains details of the method, as well as other methods 
for deriving $\boldsymbol{A}$ that we have experimented with.

\section{Results}
\label{sec:results}
We now compare results from the three methods described in the previous section.
All three methods depend on a national BHM simulation. We used a simulation that was used to produce the official UN TFR projections in WPP2012. Our version has 2,000 TFR trajectories for each country and was produced using the bayesTFR R package~\citep{Sevcikova&2011}.

For the Scale-AR(1) method, for each region $r_c$ we set the initial scaling factor to $\alpha_{r_c,P} = f_{r_c,P} / f_{c,P}$ with $P$ being the last observed time period. Then we produced projections of $\alpha_{r_c,t}$ for $t>P$ using (\ref{eq:AR1}). Finally, (\ref{eq:TFRscalesimple}) was applied, as in the case of the simple Scale method, using  each of the 2,000 TFR trajectories for country $c$ as $f_{c,t,i}$. This yielded 2,000 regional TFR trajectories.

For the one-directional BHM (1d-BHM), we ran the regional BHM while using the country posterior from the national BHM simulation.
Then we projected 2,000 regional TFR trajectories using a sample of the regional posterior parameters. We explored two versions of this model, one that accounts for correlation between regions' error terms and one that does not, the latter denoted by ``1d-BHM (indep)''.

\subsection{TFR projections}
\begin{figure}
\centering
\includegraphics[width=0.7\textwidth]{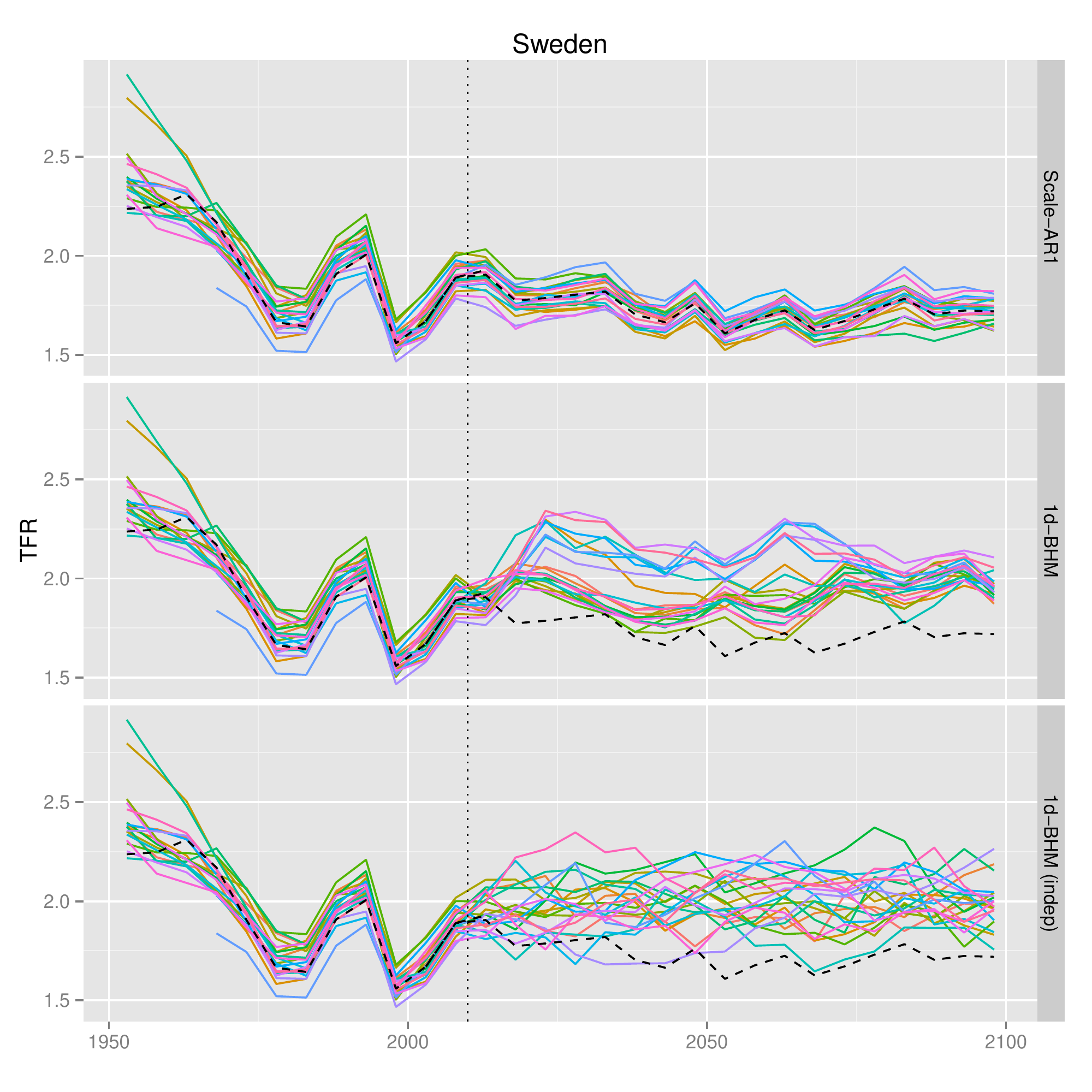}
\caption{\label{fig:1traj_sweden} Observed data and one randomly selected projection trajectory for all regions of Sweden. The projections were obtained via three different methods: Scale-AR(1) (top), the one-directional BHM that accounts for correlation (center) and the one-directional BHM that treats regions independently (bottom). The vertical dotted line marks the last observed time period. The black dashed line marks the corresponding national trajectory.}
\end{figure}

We are interested in the marginal predictive distribution of future TFR for each region. We are also interested in how reasonable the joint distributions of the trajectories between regions are. Fig.~\ref{fig:1traj_sweden} shows one randomly selected trajectory for all regions of Sweden for various methods. In the top panel the Scale-AR(1) method was used. It can be seen that all trajectories closely follow the corresponding national trajectory (black dashed line), while allowing for occasional crossovers. 
This creates a similar pattern to that seen in the observed data (to the left from the dotted vertical  line). The simple Scale method (not shown in the figure) yield trajectories perfectly parallel to the national trajectory with no crossovers.

The bottom two panels of Fig.~\ref{fig:1traj_sweden} show results from the one-directional BHM method. In the middle panel we accounted for correlation between regions, 
whereas in the bottom panel the regions' error terms were considered independent. As can be seen, this method does not yield trajectories that closely parallel the national one. Furthermore, if correlation is not taken into account, there are many more crossovers between regions than are typically seen in the past data.  

\begin{figure}
\centering
\includegraphics[width=\textwidth]{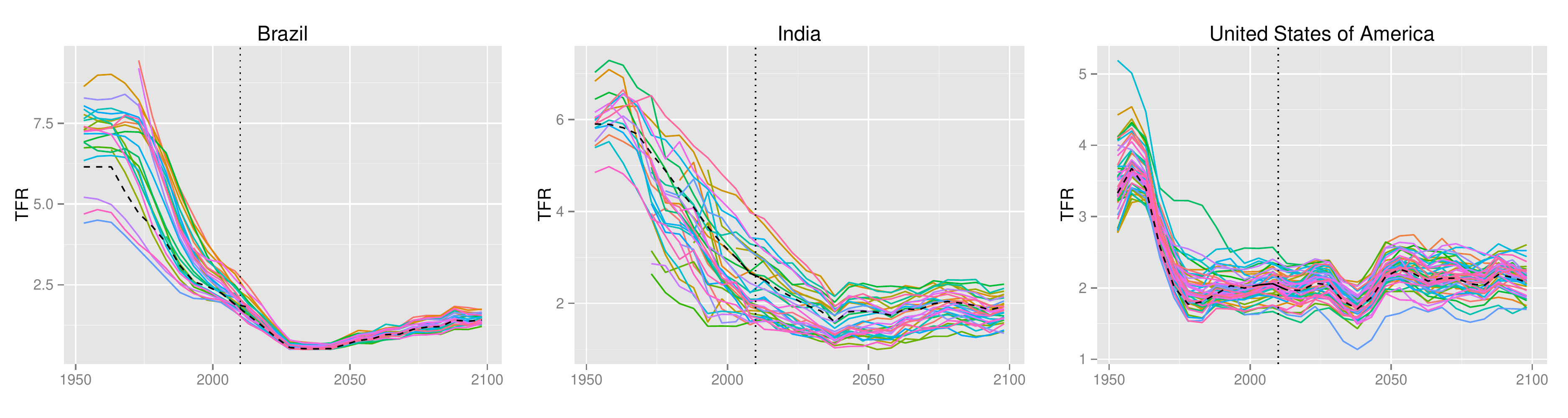}
\caption{\label{fig:1traj_selcountries} Observed data and one randomly selected projection trajectory for each region, obtained via the Scale-AR(1) method for all regions of Brazil, India and the USA. The vertical dotted line marks the last observed time period. The black dashed line marks the corresponding national trajectory.}
\end{figure}

All the 47 countries in our dataset show the same pattern in terms of the differences between the methods. In Fig.~\ref{fig:1traj_selcountries} we selected three countries for which one trajectory obtained via the Scale-AR(1) method is shown for each region (as in the top panel of Fig.~\ref{fig:1traj_sweden}). As in the case of Sweden, the trajectories are highly correlated and closely follow the national trajectory. 

\begin{figure}
\centering
\includegraphics[width=\textwidth]{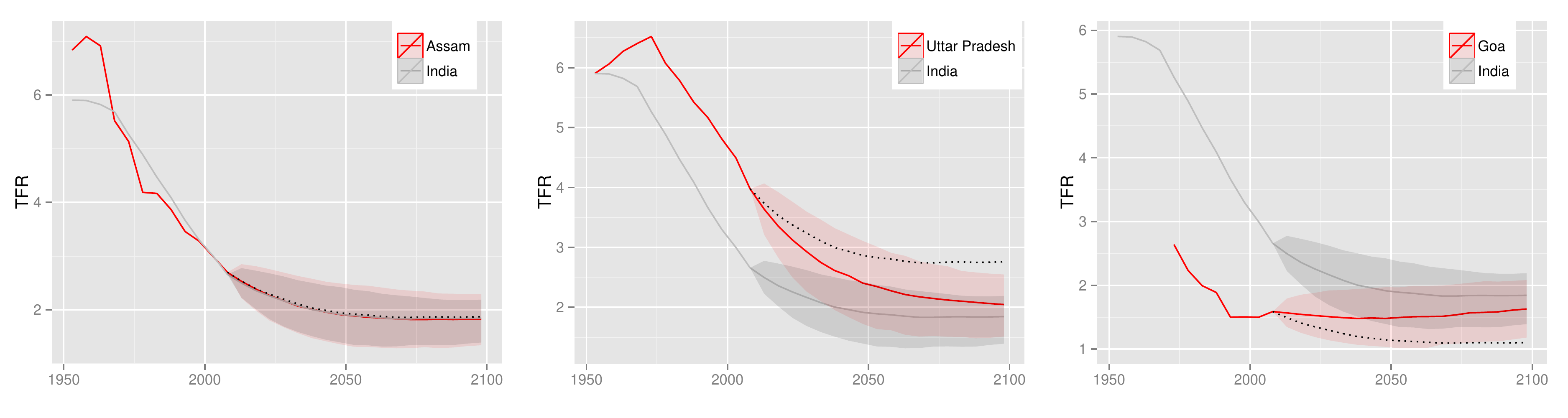}
\caption{\label{fig:predIndia} TFR projections for three regions of India: Observed data and median projections are shown by the red line, and 80\% prediction interval by the red shaded area. National data, projection median and 80\% prediction interval are shown as grey line and shaded area, respectively. The dotted line shows the median projection resulting from the simple Scale method.}
\end{figure}

In Fig.~\ref{fig:predIndia} we show the predictive median and 80\% prediction interval (red) for three regions of India from the Scale-AR(1) method (the corresponding national projection is shown in grey). They represent three different  types of regions found across all countries. The first type (in the left panel, Assam, India) is a region with a current  TFR that is very close to the national TFR. In such a case, the regional projection mostly overlaps with the national projection, with a slightly larger prediction interval. The black dotted line in the figure shows the median projection resulting from the simple Scale method. 
This would also be very close to the national median for regions of this 
type.

Uttar Pradesh in the center is a type of region where current TFR is substantially higher than the national TFR. The underlying AR(1) process causes the median projection of such a region to converge to the national median in the long term, thus decreasing the gap between them. 
If simple scaling were applied, that gap would remain constant, 
resulting in much higher projections of TFR for the region.

Finally, Goa on the right, with its current TFR well below  the national one, is projected to increase on average, again yielding a smaller gap between the national and regional medians. Here simple scaling results in much lower projections.

\subsection{Out-of-sample predictive validation}
\label{sec:validation}
We validated our methodology via predictive out-of-sample experiments, one for predicting the period 1995--2010, and one for predicting the period 1990-2010.
We first assessed the various methods in terms of average predictive performance over all regions of the 47 countries. To assess their performance for predicting aggregates (and hence, for example, in capturing the between-region correlations), we further assess the predictions of the average TFR 
across the regions of each country.

For both time periods considered, we removed the data points that corresponded to the time period to be predicted, reestimated the models, generated probabilistic projections with the various methods, and compared the projections with the observed data points. The results are shown in Table~\ref{tab:3out} for 1995-2010 and Table~\ref{tab:4out} for 1990-2010.  The measures in the left part of each table (Marginal TFR) were derived by comparing the probabilistic projections of TFR for all regions to their observed values. The quantities in the right part of the tables (Average TFR) are derived by comparing a TFR averaged over all regions of each country with the observed average TFR for each country. 

For comparison purposes, we also added the Persistence method, in which the TFR stays at the same level over time and so the forecast for all future time periods is equal to the last observed value. While this could be viewed as a straw man forecast, persistence forecasts have been found to perform surprisingly well in many forecasting contexts, and so it is worth making this comparison.

In the tables, the mean absolute error (MAE), the bias and the continuous ranked probability score (CRPS) \citep{Hersbach2000,gneiting&raftery2007} are reported.
The coverages of the 80\% and 95\% intervals are also reported.
The coverage of a prediction interval is defined as the proportion of the 
time that the truth lies in the interval. We wish the coverage to be close
to the nominal level. Thus, for example, ideally the coverage of the 80\%
interval would be close to 80\%.

\begin{table}
\caption{\label{tab:3out} Out-of-sample validation of probabilistic subnational TFR projections over 1995-2010. MAE is mean absolute error.  CRPS is continuous ranked probability score. The $80\%$ and $95\%$ columns refer to the percentage of the observations that fell within their prediction interval. The marginal TFR was validated on 3199 values;  the average TFR was validated on 137 values. The Scale-AR(1) parameters were $\phi=0.898$ and $\sigma = 0.0533$.}
\begin{center}
{\small 
\begin{tabular}{lrrrrr|rrrrr}
\hline\hline
&\multicolumn{5}{c|}{Marginal TFR} & \multicolumn{5}{c}{Average TFR} \\
\multicolumn{1}{l}{}&\multicolumn{1}{c}{MAE}&\multicolumn{1}{c}{bias}&\multicolumn{1}{c}{CRPS}&\multicolumn{1}{c}{80\%}&\multicolumn{1}{c|}{95\%}&\multicolumn{1}{c}{MAE}&\multicolumn{1}{c}{bias}&\multicolumn{1}{c}{CRPS}&\multicolumn{1}{c}{80\%}&\multicolumn{1}{c}{95\%}\tabularnewline
\hline
Scale-AR(1)&$0.205$&$-0.088$&$-0.147$&$82.0$&$96.3$&$0.172$&$-0.117$&$-0.127$&$82.5$&$95.6$\tabularnewline
1d-BHM&$0.228$&$-0.067$&$-0.167$&$75.1$&$90.1$&$0.169$&$-0.101$&$-0.123$&$71.5$&$89.1$\tabularnewline
1d-BHM (indep)&$0.228$&$-0.071$&$-0.167$&$75.2$&$89.8$&$0.169$&$-0.103$&$-0.142$&$38.7$&$50.4$\tabularnewline
Scale&$0.220$&$-0.106$&$-0.156$&$76.2$&$92.2$&$0.182$&$-0.136$&$-0.133$&$78.8$&$95.6$\tabularnewline
Persistence&$0.365$&$-0.305$&$-0.365$&\multicolumn{1}{c}{--}&\multicolumn{1}{c|}{--}&$0.334$&$-0.303$&$-0.334$&\multicolumn{1}{c}{--}&\multicolumn{1}{c}{--}\tabularnewline
\hline
\end{tabular}}
\end{center}

\end{table}

\begin{table}
\caption{\label{tab:4out} Out-of-sample validation of TFR projections over 1990-2010. The marginal TFR was validated on 4144 values;  the average TFR was validated on 180 values. The Scale-AR(1) parameters were $\phi=0.910$ and $\sigma = 0.0513$.}
\begin{center}
{\small
\begin{tabular}{lrrrrr|rrrrr}
\hline\hline
&\multicolumn{5}{c|}{Marginal TFR} & \multicolumn{5}{c}{Average TFR} \\
\multicolumn{1}{l}{}&\multicolumn{1}{c}{MAE}&\multicolumn{1}{c}{bias}&\multicolumn{1}{c}{CRPS}&\multicolumn{1}{c}{80\%}&\multicolumn{1}{c|}{95\%}&\multicolumn{1}{c}{MAE}&\multicolumn{1}{c}{bias}&\multicolumn{1}{c}{CRPS}&\multicolumn{1}{c}{80\%}&\multicolumn{1}{c}{95\%}\tabularnewline
\hline
Scale-AR(1) &$0.323$&$-0.209$&$-0.234$&$70.0$&$84.8$&$0.278$&$-0.192$&$-0.202$&$73.3$&$87.2$\tabularnewline
1d-BHM&$0.344$&$-0.207$&$-0.260$&$64.5$&$79.1$&$0.284$&$-0.187$&$-0.214$&$60.0$&$76.1$\tabularnewline
1d-BHM (indep)&$0.344$&$-0.209$&$-0.260$&$64.6$&$79.6$&$0.284$&$-0.190$&$-0.242$&$26.7$&$41.7$\tabularnewline
Scale&$0.333$&$-0.230$&$-0.245$&$65.6$&$82.0$&$0.291$&$-0.215$&$-0.210$&$72.2$&$87.2$\tabularnewline
Persistence&$0.590$&$-0.538$&$-0.590$&\multicolumn{1}{c}{--}&\multicolumn{1}{c|}{--}&$0.519$&$-0.476$&$-0.522$&\multicolumn{1}{c}{--}&\multicolumn{1}{c}{--}\tabularnewline
\hline
\end{tabular}}
\end{center}

\end{table}

The appendix gives details of the derivation of these metrics. For MAE and bias, the smaller the absolute value the better. For the two coverage columns an ideal method would match the numbers to the corresponding percentage. The CRPS  is a combination of an error-based and a variation-based measure, and thus we give it a high weight when selecting the best method. In this case, a better method corresponds to a larger value of CRPS. 

For the marginal TFR, the Scale-AR(1) method performs best in terms of CRPS, MAE and coverage. The simple Scale method comes in second. However, we would not recommend using the simple Scale method because it produces trajectories 
that are unrealistic in that they do not allow the possibility of crossovers
between regions, as mentioned previously.
Note that by design, the Scale-AR(1) method yields larger uncertainty than the simple Scale method, which in this case translates to a better coverage and CRPS. The Scale method includes only the uncertainty from the national BHM model, whereas the Scale-AR(1) method has in addition the uncertainty included in the AR(1) process. There is essentially no difference between the 1d-BHM with and without correlation for the marginal TFR. This is expected, as the correlation plays a role only in aggregated indicators.

For the average TFR, the Scale-AR(1) and 1d-BHM have similar performance in terms of CRPS (one is better in Table~\ref{tab:3out}, the other in Table~\ref{tab:4out}). However, Scale-AR(1) has consistently better coverage. Here we see a big difference in coverage between the two versions of 1d-BHM, which does not have good performance if correlation between regions is not taken into account.
The good performance of the Scale-AR(1) method suggests that it is accounting adequately for between-region spatial correlation.

\section{Discussion}
\label{sec:discussion}
We have developed several methods for subnational probabilistic 
projection of TFR, and applied them to data from 47 very diverse countries.
All the methods take the national projections from the UN method as their
starting point. We found that all the methods we propose performed well in terms
of out-of-sample predictive performance, and outperformed
a simple baseline persistence method.

In the best method, the national trajectories are scaled by a 
region-specific scaling factor which itself is allowed to 
vary stochastically but slowly over time. 
One competing method treats the regions in 
the same way as countries are treated in the UN's BHM, but this does
not yield enough within-country correlation. Even when we 
introduce between-region correlation into this model, 
it still does not have enough within-country correlation overall.

We have compared several different methods, but there are still others
in the literature. \citet{Rayer&2009} considered ex-post assessment
of predictive uncertainty for U.S. counties, extending the
national ex-post approach of \citet{Keyfitz1981} and \citet{Stoto1983}
to the subnational context. \citet{Raymer&2012} used a vector autoregressive
model for crude birth rates in three regions of England. While these
methods may work well for developed countries that have had low fertility
for an extended period, they do not capture the systematic variation
in fertility decline rates among higher-fertility countries
documented by \citet{Alkema&2011}, and so they may not be so appropriate
for our goal here, of developing a method applicable to countries at all
levels of the fertility transition.

The extant method closest to our preferred Scale-AR(1) method is one proposed
by \citet{Wilson2013}, who also proposed scaling a national TFR forecast
by a region-specific scale factor that varies according to an AR(1) model,
and applied it to Sidney, Australia. However, there are several differences
between the Scale-AR(1) method we propose here, and Wilson's approach for TFR.
The national TFR forecast used by Wilson is based on an AR(1) process
centered around an externally-specified main forecast. As discussed, this
may not carry over well to higher-fertility countries. Our method, in
contrast, is centered around the probabilistic forecast from the UN's BHM,
which is designed to work well for countries at all fertility levels and includes uncertainty about national projections.
Also, in our method the model is statistically estimated, while in
Wilson's approach the parameters are adjusted manually. 

Our preferred Scale-AR(1) method does not incorporate spatially-indexed
between-region correlation. Instead, spatial correlation is modeled
by a strong country effect. Our 1-d BHM method did incorporate spatial
correlation in the variant that includes between-region correlation
estimated from the data (especially methods 8--11 described
in the Appendix section on estimating the error correlations). However,
this did not allow us to include enough between-region correlation.
This may be because within-country correlation seems to be dominated
by a strong country effect rather than spatially indexed correlation,
as can be seen for example for Sweden in Fig.~\ref{fig:1traj_sweden}.
This is also shown by the good calibration of the Scale-AR(1).
Thus we feel it is likely that adding additional spatial correlation
would not substantially improve fit of the model to the data at hand.

In addition to providing guidance for subnational projections,
our results give insight into how subnational fertility evolves
in a modern context. They suggest that there is substantial within-country
correlation and convergence. This confirms the observations and hypotheses
of Watkins (1990, 1991) for Europe to 1960. 
It further extends them from just Europe to a range
of countries from around the world, and indicates that, broadly speaking,
similar patterns continue to hold a half-century later.

\paragraph{Acknowledgements:} This work was supported by NICHD grants
R01 HD054511 and R01 HD070936.


\appendix{}
\section{Appendix: Methods}
\subsection{Estimation of the Scale-AR(1) parameters}
\label{app:ar1}
Here we give details on estimating parameters of the Scale-AR(1) model.

The model is based on an AR(1) process for region-specific
scale factors $\alpha_{r_c,t}$ centered at one, namely
\begin{equation}
\label{eq:appAR1}
\alpha_{r_c,t} - 1 =  \phi(\alpha_{r_c,t-1}-1)  + \varepsilon_{r_c,t}, \quad \text{with} \;\; \varepsilon_{r_c,t} \stackrel{iid}{\sim}  N(0, \sigma_{c}^2) .
\end{equation}
We impose the restriction that the scale factors not diverge indefinitely
over time. We implement this by requiring that $\sigma^2_{c}$ is such that 
\begin{equation}
\label{eq:appAR1restr}
\lim_{t \rightarrow \infty}  \text{Var}(\alpha_{r_c,t}) \leq \text{Var}_{q \in R_c}(\alpha_{q,t=P}) ,
\end{equation}
where $P$ denotes the present time period and $R_c$ denotes the set of regions in country $c$. This yields
\begin{equation}
\label{eq:appsigmac}
\sigma^2_{c}= \min\{ \sigma^2, (1-\phi^2)\text{Var}_{r \in R_c}(\alpha_{r,t=P})\} 
\end{equation}
We are interested in estimating the country- and region-independent parameters  $\phi$ and $\sigma$. We know from the observed data that the standard deviation of $\alpha_{r_c,t}$ declines as TFR declines, which is also in line with
the theoretical expectations of Watkins (1990, 1991).
\nocite{Watkins1990,Watkins1991}
Thus we need to find asymptotic values for those parameters.

Let $\Delta \alpha_{r_c,t}$ denote the first order differences over time, 
namely
\begin{equation}
\label{eq:appAR1dif}
\Delta \alpha_{r_c,t} = \alpha_{r_c,t} - \alpha_{r_c,t-1} .
\end{equation}
Then 
\begin{eqnarray}
\label{eq:appAR1lim1}
\lim_{t \rightarrow \infty}  \text{Var}(\alpha_{r_c,t}) &=& \frac{\sigma^2}{1-\phi^2}  \quad\quad \text{and} \\
\label{eq:appAR1lim2}
\lim_{t \rightarrow \infty}  \text{Var}(\Delta \alpha_{r_c,t}) &=& 2(1-\phi)\text{Var}(\alpha_{r_c,t}) .
\end{eqnarray}
Equations~(\ref{eq:appAR1lim1}) and~(\ref{eq:appAR1lim2}) imply that 
 \begin{eqnarray}
\label{eq:appAR1phi}
\phi &=& 1- \frac{\text{Var}(\Delta \alpha_{r_c,t})}{2 \text{Var}(\alpha_{r_c,t})}  \quad\quad \text{and}  \\
\label{eq:appAR1sigma}
\sigma^2 &=& \text{Var}(\Delta \alpha_{r_c,t}) - (1-\phi)^2\text{Var}(\alpha_{r_c,t}) .
\end{eqnarray}

\begin{figure}
\centering
\includegraphics[width=\textwidth]{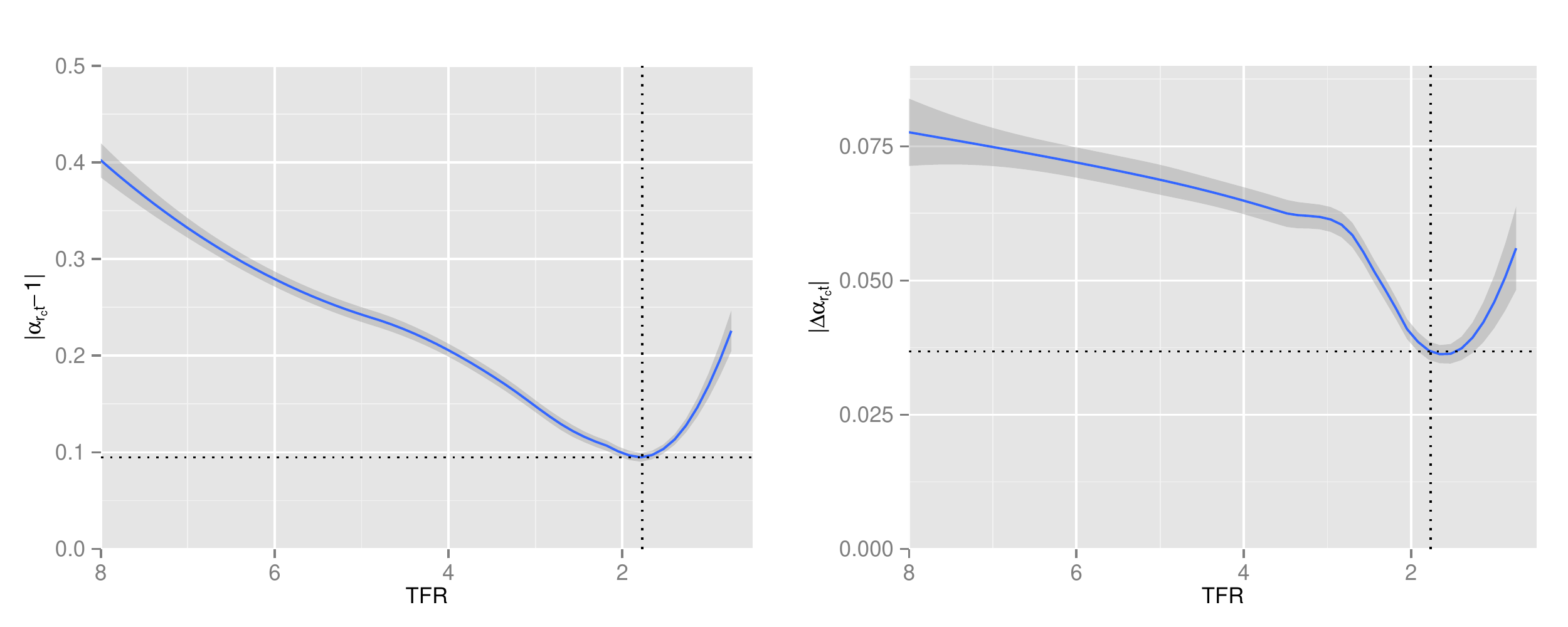}
\caption{\label{fig:app-alpha-loess} The loess curve for $|\alpha_{r_c,t}-1| \sim \textrm{TFR}$ (left panel) and for  $| \Delta \alpha_{r_c,t}| \sim \textrm{TFR}$ (right panel). They are based on $9,566$ data points.}
\end{figure}

Assuming a normal distribution of $\alpha_{r_c,t}$ we can write
\begin{eqnarray}
\label{eq:appAR1var1}
 \text{Var}(\alpha_{r_c,t}) &=& \frac{\pi}{2}(E[ | \alpha_{r_c,t}-1|])^2 , \\
\label{eq:appAR1var2}
\text{Var}(\Delta \alpha_{r_c,t}) &=&  \frac{\pi}{2}(E[ | \Delta \alpha_{r_c,t}|])^2 .
\end{eqnarray}
From the observed data we know  that both $|\alpha_{r_c,t}-1|$ and  $| \Delta \alpha_{r_c,t}|$ decline as TFR declines (see Fig.~\ref{fig:app-alpha-loess}). 
The nonparametrically estimated conditional expectation of
$|\alpha_{r_c,t}-1|$ given TFR reaches a minimum, as a function of TFR,
of $0.09475$ at TFR $=1.768$, as shown by the  dotted lines in
Fig.~\ref{fig:app-alpha-loess}.
At this level of TFR, the nonparametrically estimated value of 
$E (| \Delta \alpha_{r_c,t}|)$ is  0.03678, which is close to its minimum. 
Taking the mean of $\alpha_{r_c,t}$ to be 1, and using the fact that 
the standard deviation of a normal random variable is $\sqrt{\pi/2}$ times
its mean absolute deviation, we find that
 \begin{eqnarray}
 \label{eq:appAR1sd1}
 \text{SD}( \alpha_{r_c,t}) & = & \sqrt{\frac{\pi}{2}} 0.09475 = 0.11875 , \\
 \label{eq:appAR1sd2}
 \text{SD}(  \Delta \alpha_{r_c,t}) & = & \sqrt{\frac{\pi}{2}} 0.03678 = 0.04610 .
\end{eqnarray}

Substituting the values from Equations (\ref{eq:appAR1sd1}) 
and (\ref{eq:appAR1sd2}) for Var$(\alpha_{r_c,t})$ and Var$(\Delta \alpha_{r_c,t})$  into Equations (\ref{eq:appAR1phi}) and (\ref{eq:appAR1sigma}) gives 
\begin{eqnarray}
\label{eq:appAR1est}
\phi &=& 0.92464 , \\
\sigma&=& 0.04522.  \nonumber
\end{eqnarray}
These are the values we use for our projections.

\subsection{Estimating the Error Correlations}
\label{app:correlation}
%
The model errors are defined by
\begin{equation*}
\boldsymbol {\varepsilon_t} \sim N(0, \Sigma_t = \boldsymbol {\sigma'_t} \boldsymbol{A} \boldsymbol {\sigma_t}) .
\end{equation*}
We have experimented with eleven different ways of estimating the correlations of the errors, which are the elements $A_{rs}$ of the matrix~$\boldsymbol{A}$. Let $e_{r,t}$ denote the model error of region $r$ at time $t$. 
Let  $\boldsymbol{\tilde{A}}$ denote  a matrix where each element $\tilde{A}_{rs}$ is the empirical correlation between $e_{r\cdot}$ and $e_{s\cdot}$ over all time periods~$t$, namely
\begin{equation}  
 \tilde{A}_{r,s} = \frac{\sum_T e_{r,t} e_{s,t}}{\sqrt{\sum_T e_{r,t}^2}\cdot\sqrt{\sum_T e_{s,t}^2}} .
 \end{equation}
Furthermore, $\boldsymbol{\tilde{A}}$ truncated at zero will be denoted by $\boldsymbol{\tilde{A}_{[\geq 0]}}$. If positive definiteness is assured, 
it is denoted by $\boldsymbol{\tilde{A}^*}$. 

We considered the following methods for estimating the matrix $\boldsymbol{A}$.
In the first seven methods,
all the within-country correlations are taken to be equal.
The estimator of $\boldsymbol{A}$ is denoted by $\hat{\boldsymbol{A}}$.
In all cases, $\hat{A}_{r,r}=1$ for all $r$, so in what follows,
$\hat{A}_{r,s}$ refers to the cases where $r \neq s$.
\begin{enumerate}
\item $\hat{A}_{r,s} = \text{mean}\{a_{ij} \in \boldsymbol{\tilde{A}}_{[\geq 0]} \text{ and } i\neq j\}$ for all $r \neq  s$. 
\item $\hat{A}_{r,s} = \text{median}\{a_{ij} \in \boldsymbol{\tilde{A}}_{[\geq 0]} \text{ and } i\neq j\}$ for all $r \neq  s$.
\item $\hat{A}_{r,s}$ is the Bayesian posterior mean of the intraclass correlation coefficient. 
\item $\hat{A}_{r,s}$ is the Bayesian posterior mode of the intraclass correlation coefficient. 
\item Similar to 3., but with errors divided by $\sqrt{1/n \sum_{r,t} e_{r,t}^2}$ with $n$ being the number of available errors. 
\item Similar to 4., but with errors divided by $\sqrt{1/n \sum_{r,t} e_{r,t}^2}$ with $n$ being the number of available errors. 
\item $\hat{A}_{r,s}=B/C$ for all $r \neq  s$, where
\begin{eqnarray*}
B & = & \frac{1}{N_b}\sum_{t=1}^T \sum_{i=1}^{R-1} \sum_{j=i+1}^R e_{i,t} \cdot e_{j,t} , \\
C & = & \frac{1}{N_c}\sum_{t=1}^T \sum_{i=1}^R e_{i,t}^2 ,
\end{eqnarray*}
$N_b$ and $N_c$ the number of  terms in the corresponding sum that are not missing, and $R$ is the number of regions.
\item The estimator of $\boldsymbol{A}$ is an approximation to the elementwise posterior median with uniform prior $U[0,1]$, namely: 
\[
\hat{\boldsymbol{A}} = \frac{\bar{T}-1}{\bar{T}} \boldsymbol{\tilde{A}^*}_{[\geq 0]} + \frac{1}{2\bar{T}} ,
\]
  where $\bar{T}$ is the average number of time periods per region. It is a weighted average of the prior mean and the data.
Note that this is our chosen method. 
  
  We will now show that if $\boldsymbol{\tilde{A}^*}_{[\geq 0]}$ is positive definite, then $\boldsymbol{A}$ is also positive definite. We can write
\begin{equation}  
   \hat{\boldsymbol{A}} =  \frac{T-1}{T}  \boldsymbol{B} + \frac{1}{2T} \boldsymbol{J},              
\end{equation}
where $\boldsymbol{B}$ is positive definite and $\boldsymbol{J}$ is the matrix all of whose entries are 1.

Now  $\hat{\boldsymbol{A}}$ is positive definite if and only if $x' \hat{\boldsymbol{A}}x > 0$ for all $x \neq 0$. Now
\begin{equation} 
\label{eq:proof-PD2}
x'\hat{\boldsymbol{A}}x =  \frac{T-1}{T} x'\boldsymbol{B} x +  \frac{1}{2T} x'\boldsymbol{J}x.    
\end{equation}
The first term on the right-hand side of (\ref{eq:proof-PD2}) is positive by definition,
since $\boldsymbol{B}$ is positive definitive. The second term is non-negative:
\begin{equation} 
x'\boldsymbol{J}x = (\sum_{i=1}^n x_i)^2 \geq 0.
\end{equation}
Thus (\ref{eq:proof-PD2}) is positive and so $\hat{\boldsymbol{A}}$ is positive definite. 
 \item Similar to 8. but with elements of $\boldsymbol{\tilde{A}}$ computed as 
 \[
 \tilde{A}_{r,s} = \frac{1/T\sum_T e_{r,t} e_{s,t}}{\sqrt{1/T\sum_T e_{r,t}^2}\cdot\sqrt{1/T\sum_T e_{s,t}^2}}
 \]
 \item Bayesian method introduced by \citet{FosdickRaftery2014}: 
 First, standardize $e_{r,t}$ by dividing the errors by $\sqrt{1/n \sum_{r,t} e_{r,t}^2}$ with $n$ being the number of available errors. Then the 
 elements of the estimated correlation matrix $\hat{\boldsymbol{A}}$ 
are given as 
 \[
 \hat{A}_{r,s} =  \frac{\int_0^1 \rho \left(\frac{1}{\sqrt{1-\rho^2}}\right)^T \exp \left[ -\frac{1}{2(1-\rho^2)}\left[ SS_r -2\rho SS_{r,s} + SS_s\right] \right] d\rho}{\int_0^1 \left(\frac{1}{\sqrt{1-\rho^2}}\right)^T \exp \left[ -\frac{1}{2(1-\rho^2)}\left[ SS_r -2\rho SS_{r,s} + SS_s\right] \right] d\rho} ,
 \]
 where $SS_r = \sum_T e_{r,t}^2$, $SS_s = \sum_T e_{s,t}^2$, and $SS_{r,s} = \sum_T e_{r,t}e_{s,t}$. Note that we are summing only over those time periods for which both countries, $r$ and $s$, have errors available. 
 \item Similar to 10., but using $(T+1)$ instead of $T$ in both the nominator and the denominator. This corresponds to the arcsin prior in \citet{FosdickRaftery2014}. Note that a version of this method was tested where the errors were not standardized, but it performed less well, producing smaller correlations.
\end{enumerate}

\cite{FosdickRaftery2014} found that correlations between countries were
quite different for high and low TFR values. In light of this, we 
estimated two separate correlation matrices, one for the cases where
the country had overall TFR 5 or above, and the other when the TFR was 
below 5.

The estimated correlation matrices resulting from methods 1.-7.  have the same value for all off-diagonal elements. The elements of matrices resulting from methods 8.-11. differ from one another. In the latter case, all non-defined elements are set to the mean of the off-diagonal elements.

\subsection{Out of Sample Validation Measures} 
\label{app:out-of-sample}
This section provides detailed definitions for our out of sample validation measures. 

We denote by $C$ the number of countries in our dataset, by $R_c$ the number of regions for country $c$, by $R$ the total number of regions, so that $R=\sum_{c=1}^C R_c$, and by $T$ the number of time periods over which we validate. Furthermore, $f_{r_c,t}$ denotes the observed TFR, and $\hat{f}_{r_c,t}$ denotes the point projection of the TFR (median of the predictive distribution), respectively,  for region $r$ of country $c$ at time $t$.

The mean absolute error, MAE, is given by
\begin{equation}
\textrm{MAE}=\frac{1}{RT} \sum_{c=1}^C \sum_{r=1}^{R_c} \sum_{t=1}^T |f_{r_c,t} - \hat{f}_{r_c,t}| \,.
\end{equation}

The bias is given by
\begin{equation}
\textrm{bias}=\frac{1}{RT} \sum_{c=1}^C \sum_{r=1}^{R_c} \sum_{t=1}^T (f_{r_c,t} - \hat{f}_{r_c,t}) \,.
\end{equation}

The Continuous Ranked Probability Score (CRPS) is an overall measure of the quality of a probabilistic forecast. If we are predicting the quantity $X$
(here the TFR), and produce the predictive distribution $F$, 
and observe a value $x$, then the CRPS is defined by:
\[
\text{CRPS}(F;x) = \frac{1}{2} E_F |X'-X''| - E_F |X-x| ,
\]
where $X'$ and $X''$ are independent copies of a random variable with the 
distribution $F$ (\citet{gneiting&raftery2007}, Eq. 21).
We average the resulting values of CRPS across observations.
Note that for the persistence method, the first part of the equation is zero.  
It is not simple to calculate the expectation, $E_F$, under the distribution $F$
analytically, so we did it by simulation. To obtain the expectation $E_F$, 
we sampled 5,000 values at random from the distribution $F$ and took
the average of the corresponding predictands.

So far we have compared the methods for the 
TFR for all regions and for the average
TFR for a country, taking the unweighted average over all regions 
in the country (Tables~\ref{tab:3out} and~\ref{tab:4out}). Here, we add a comparison of the various correlation methods discussed in Appendix~\ref{app:correlation} for the average TFR (Tables~\ref{tab:3out-avg} and \ref{tab:4out-avg}). The first row shows results when no correlation is taken into account. The number in parentheses of the following rows corresponds to the numbering of the eleven methods in the Appendix.
In our study, we used method 8.

\begin{table}
\caption{\label{tab:3out-avg} Out-of-sample validation of average TFR projections over 1995-2010, validated on 137 values.}
\begin{center}
\begin{tabular}{lrrrrr}
\hline\hline
\multicolumn{1}{l}{}&\multicolumn{1}{c}{MAE}&\multicolumn{1}{c}{bias}&\multicolumn{1}{c}{CRPS}&\multicolumn{1}{c}{80\%}&\multicolumn{1}{c}{95\%}\tabularnewline
\hline
1d-BHM (indep.)&$0.169$&$-0.103$&$-0.142$&$38.7$&$50.4$\tabularnewline
1d-BHM (1.)&$0.170$&$-0.101$&$-0.124$&$69.3$&$89.1$\tabularnewline
1d-BHM (2.)&$0.167$&$-0.102$&$-0.122$&$74.5$&$89.8$\tabularnewline
1d-BHM (3.)&$0.168$&$-0.104$&$-0.127$&$62.0$&$80.3$\tabularnewline
1d-BHM (4.)&$0.167$&$-0.105$&$-0.127$&$62.8$&$81.0$\tabularnewline
1d-BHM (5.)&$0.167$&$-0.106$&$-0.124$&$69.3$&$84.7$\tabularnewline
1d-BHM (6.)&$0.167$&$-0.106$&$-0.125$&$68.6$&$83.9$\tabularnewline
1d-BHM (7.)&$0.168$&$-0.104$&$-0.125$&$65.7$&$83.2$\tabularnewline
1d-BHM (8.)&$0.169$&$-0.101$&$-0.123$&$71.5$&$89.1$\tabularnewline
1d-BHM (9.)&$0.169$&$-0.100$&$-0.123$&$70.1$&$89.1$\tabularnewline
1d-BHM (10.)&$0.168$&$-0.102$&$-0.123$&$70.8$&$89.1$\tabularnewline
1d-BHM (11.)&$0.167$&$-0.103$&$-0.122$&$71.5$&$89.8$\tabularnewline\hline
Scale-AR(1)&$0.172$&$-0.117$&$-0.127$&$82.5$&$95.6$\tabularnewline
Scale&$0.182$&$-0.136$&$-0.133$&$78.8$&$95.6$\tabularnewline
Persistence&$0.334$&$-0.303$&$-0.334$&\multicolumn{1}{c}{--}&\multicolumn{1}{c}{--}\tabularnewline
\hline
\end{tabular}\end{center}

\end{table}

\begin{table}
\caption{\label{tab:4out-avg} Out-of-sample validation of average TFR projections over 1990-2010, validated on 180 values.}
\begin{center}
\begin{tabular}{lrrrrr}
\hline\hline
\multicolumn{1}{l}{}&\multicolumn{1}{c}{MAE}&\multicolumn{1}{c}{bias}&\multicolumn{1}{c}{CRPS}&\multicolumn{1}{c}{80\%}&\multicolumn{1}{c}{95\%}\tabularnewline
\hline
1d-BHM (indep.)&$0.284$&$-0.190$&$-0.242$&$26.7$&$41.7$\tabularnewline
1d-BHM (1.)&$0.284$&$-0.187$&$-0.213$&$60.0$&$77.2$\tabularnewline
1d-BHM (2.)&$0.279$&$-0.190$&$-0.210$&$63.3$&$78.3$\tabularnewline
1d-BHM (3.)&$0.284$&$-0.192$&$-0.217$&$55.6$&$71.1$\tabularnewline
1d-BHM (4.)&$0.283$&$-0.192$&$-0.216$&$55.6$&$70.0$\tabularnewline
1d-BHM (5.)&$0.284$&$-0.194$&$-0.215$&$57.8$&$76.1$\tabularnewline
1d-BHM (6.)&$0.284$&$-0.194$&$-0.216$&$57.8$&$74.4$\tabularnewline
1d-BHM (7.)&$0.284$&$-0.191$&$-0.214$&$56.7$&$72.2$\tabularnewline
1d-BHM (8.)&$0.284$&$-0.187$&$-0.214$&$60.0$&$76.1$\tabularnewline
1d-BHM (9.)&$0.284$&$-0.187$&$-0.213$&$60.0$&$76.1$\tabularnewline
1d-BHM (10.)&$0.285$&$-0.188$&$-0.215$&$60.0$&$76.7$\tabularnewline
1d-BHM (11.)&$0.284$&$-0.189$&$-0.214$&$61.1$&$77.2$\tabularnewline\hline
Scale-AR(1)&$0.278$&$-0.192$&$-0.202$&$73.3$&$87.2$\tabularnewline
Scale&$0.291$&$-0.215$&$-0.210$&$72.2$&$87.2$\tabularnewline
Persistence&$0.519$&$-0.476$&$-0.522$&\multicolumn{1}{c}{--}&\multicolumn{1}{c}{--}\tabularnewline
\hline
\end{tabular}\end{center}

\end{table}

\afterpage{%
    \clearpage
    \thispagestyle{empty}
    \begin{landscape}
    \centering
        {\tiny
\begin{longtable}{llrrp{17cm}}
\caption{Sources of the data used in the study.} \\ 
  \hline
Country & Geographic.units & Nunits & Nobs. & Subnational.TFR.data.source \\ 
  \hline
Argentina & Provinces (Jurid.) &  24 & 321 & Pantelides, Edith Alejandra (2006). La Transiciﾗn de la fecundidad en la Argentina 1869-1947. CENEP, Centro de Estudios de Poblacion, Cuaderno del CENEP No. 54 ; Pantelides, Edith Alejandra (1989). La Fecundidad Argentina desde Mediados del Siglo XX. CENEP, Centro de Estudios de Poblacion, Cuaderno del CENEP No. 41 ; Instituto Nacional de Estadﾒstica y Censo, INDEC (2012). Dinﾇmica y estructura de la poblaciﾗn. Tasa bruta de natalidad por provincia. Aﾖos 1980 - 2009 (http://www.indec.gov.ar/principal.asp?id\_tema=7924) \\ 
  Brazil & States &  27 & 332 & IBEGI (2012). 1940-2010 Censuses (P/F ratio adjusted) and 2001-2009 PNAD surveys \\ 
  Canada & Provinces-Territ. &  13 & 189 & Statistics Canada, Vital Statistics and Demography Division (2012). Fertility rate by age of mother, 1921 to 2009 \\ 
  Chile & Regions &  15 & 136 & CEPAL/CELADE Redatam+SP 10/26/2012. Estimaciﾗn Indirecta de la Fecundidad. Chile - Censo de Poblaciﾗn y Vivienda 1982-2002 ; National Statistical Institute, Chile (INE) 1997-2010 estimates (http://www.ine.cl/) \\ 
  Costa Rica & Provinces &   7 &  77 & Rosero-Bixby, L. (2012). Unpublished Estimates of fertility in the provinces of Costa Rica 1956-2011.  Centro Centroamericano de Poblaciﾗn (CCP) of the Universidad de Costa Rica (UCR). The estimates are based in the Vital Statistics on Births, published by the Instituto Nacional de Estadﾒstica y Censos (INEC) and interpolation of the female population from the 1950, 1963, 1973, 1984, 2000 and 2011 census.   Both births and population were corrected following: Brenes, G. (2012). Evaluation of the 2011 census and demographic estimates for the period 1950-2011.  Unpublished manuscript, CCP-UCR-INEC. \\ 
  Cuba & Provinces &  15 & 120 & Oficina Nacional de Estad�sticas de Cuba. �Series Demogr�ficas 1982 � 2002�, Tomo I . 3. Tasa de Natalidad (por 1000 habitantes) seg�n provincia de residencia, per�odo 1970-2002 ; Oficina Nacional de Estad�sticas de Cuba. Anuario Demogr�fico de Cuba 2005-2011. '3.14 - Tasas del movimiento natural  por provincias  \\ 
  Ecuador & Provinces &  22 & 198 & CEPAL/CELADE Redatam+SP 10/26/2012. Estimaciﾗn Indirecta de la Fecundidad. Ecuador - Censo de Poblaciﾗn y Vivienda 1982-2010 \\ 
  Mexico & States &  32 & 256 & Partida V, (2008). Proyecciones de la Poblaciﾗn de Mﾎxico, de las Entidades Federativas, de los Municipios y de las Localidades 2005-2050 Mﾎxico, D.F.: CONAPO. Courtesy of Fatima Juarez. ; CEPAL/CELADE Redatam+SP 10/19/2012. Estimaciﾗn Indirecta de la Fecundidad. Mﾎxico - Censo de Poblaciﾗn y Vivienda 2010. \\ 
  Panama & Provinces &  12 &  90 & CEPAL/CELADE Redatam+SP 10/26/2012. Estimaciﾗn Indirecta de la Fecundidad. Panama - Censo de Poblaciﾗn y Vivienda 1990-2010 \\ 
  Paraguay & Departments &  18 & 144 & CEPAL/CELADE Redatam+SP 10/26/2012. Estimaciﾗn Indirecta de la Fecundidad. Paraguay - Censo de Poblaciﾗn y Vivienda 1982-2002 \\ 
  USA & States &  51 & 1184 & US Census Office (1902). 1900 Census Reports, Vol. III, Vital statistics, Part I ; US National Office of Vital Statistics (1947). Vital statistics rates in the United States 1900-1940 ; National Center for Health Statistics, NCHS: Birth and Fertility Rates for States and Metropolitan Areas United States for States and Metropolitan Areas United States DHEW Publication No. (HRA) 78-1905 and  http://www.cdc.gov/nchs/data\_access/vitalstats/VitalStats\_Births.htm \\ 
  Uruguay & Departments &  19 & 171 & CEPAL/CELADE Redatam+SP 10/19/2012. Estimaciﾗn Indirecta de la Fecundidad. Uruguay Censo de Poblaciﾗn y Vivienda. 1985-2011. \\ 
  Venezuela & States &  25 & 196 & CEPAL/CELADE Redatam+SP 10/19/2012. Estimaciﾗn Indirecta de la Fecundidad. Venezuela - Censo de Poblaciﾗn y Vivienda 1990-2011 \\ 
  Australia & States $\backslash$\& Territ. &   8 & 104 & Australian Bureau of Statistics, Australian Historical Population Statistics, 2008; Australian Demographic Statistics, June 2010. \\ 
  China & Provinces &  31 & 329 & Yao, X. (1995). China ferility data collection. China Population Press: Beijing ; 1982-2010 censuses \\ 
  China, Shanghai & Districts &  20 & 100 & Shanghai population data (de jure population only). http://rkjsw.sh.gov.cn/stat/ \\ 
  India & States &  33 & 510 & Ram, U. and F. Ram 2009. �Fertility in India: Policy Issues and Program Challenges,� in K.K. Singh, R.C. Yadava and Arvind Pandey (eds.) Population, Poverty and Health: Analytical Approaches. Hindustan Publishing Company, New Delhi India. ; Rele, J.R. 1987. �Fertility Levels and Trends in India, 1951-81,� Population and Development Review, 13, 3, 513-530 ; Office of registar General of India 1971-2010 Sample Registration System \\ 
  Indonesia & Provinces &  33 & 236 & ESCAP (1987). Levels and trends of fertility in Indonesia based on the 1971 and 1980 population censuses - a study on regional differentials. ; 1971-1990 Population Censuses, 1985 Intercensal Population Surveys, 1991 and 1994 Indonesia DHS, 2010 Census Own-Children \& Rele fertility estimates - Indonesia: Fertilitas Penduduk Indonesia - Hasil Sensus Penouduk 2010 (http://sp2010.bps.go.id/files/ebook/fertilitas\%20penduduk\%20indonesia/index.html) \\ 
  Iran & Provinces &  26 & 152 & Statistical Centre of Iran (2001). Estimation of Levels and Patterns of Fertility in Iran - With Application of Own-Children Method (1972 � 1996): Table 18, 21 and 22 ; Abbasi-Shavazi M.J. and P. McDonald (2005). National and Provincial-Level Fertility Trends in Iran, 1972-2000 [ANU-WP94] \\ 
  Japan & Prefectures &  47 & 752 & Ministry of Health, Labour and Welfare, National Institute of Population and Social Security Research, 2010 Demographic Sourcebook: Vital Statistics ; 1960-2010 estimates (http://www.e-stat.go.jp) \\ 
  Korea & Provinces &  16 & 135 & Rele J.R. (1988). 70 Years of Fertility Change in Korea: New Estimates from 1916 to 1985, Asia-Pacific Population Journal, Vol. 3, No. 2 ; Korea National Statistical Office (2004). The Population of Korea. Table 3.1 ; Statistics Korea KOSIS 1970-2011 Statistics Table for Live Births, Crude Birth Rate and Total Fertility Rate by Province \\ 
  Thailand & Regions &   5 &  44 & Pejaranonda C. (1985). Declines in fertility by district in Thailand : An analysis of the 1980 census. Bangkok, Thailand, United Nations. Economic and Social Commission for Asia and the Pacific (ESCAP), (Asian Population Series No. 62-A)) ; National Statistics Office (1997). Report on the 1995-1996 Survey of population change. UNFPA (2011). Impact of demographic change in Thailand. \\ 
  Turkey & Provinces &  81 & 810 & National Academy of Sciences, Committee on Population and Demography. 1982. Trends in Fertility and Mortality in Turkey, 1935-1975, by Frederic C. Shorter, Miroslav Macura, and the Panel on Turkey. Report No. 8. Provincial estimates courtesy of Prof. Sinan Turkyilmaz (Hacettepe University Institute of Population Studies) for 1998, 2003 and 2008 TDHS  \\ 
  Austria & States &   9 &  90 & Statistics Austria (http://www.statistik.at) \\ 
  Belgium & Regions &   3 &  24 & Statistics Belgium (2012). NI\_03.24\_historique\_FR\_v4: Taux de fﾎconditﾎ selon l'ﾉge des femmes atteint dans l'annﾎe, de 15 ﾈ 49 ans pour le pays et les regions \\ 
  Bulgaria & Regions &   6 &  24 & EuroStat (2012). Fertility rates by age and NUTS 2 regions, demo\_r\_frate2 (http://ec.europa.eu/eurostat/data/database) and National Statistical institute (http://www.nsi.bg/en) \\ 
  Czech Republic & Regions &  14 &  84 & Czech Statistical Office (2012). Demographic Yearbooks 1982-1990 (http://www.czso.cz/csu/redakce.nsf/i/casova\_rada\_demografie) ; Fertility in Czech republic since 1991-2006 (national and regional level - NUTS3 and LAU1) ; Demographic Yearbook of the Regions of the Czech republic 2002-2011 (http://www.czso.cz/csu/2012edicniplan.nsf/engpubl/4027-12-eng\_r\_2012) \\ 
  Denmark & Counties &  16 &  96 & Statistics Denmark (2012). Statbank (http://www.statbank.dk/statbank5a/default.asp?w=1280) \\ 
  Estonia & Counties &  16 &  64 & Statistics Estonia (2012). http://pub.stat.ee/px-web.2001/I\_Databas/Population/01Population\_indicators\_and\_composition/02Main\_demographic\_indicators/02Main\_demographic\_indicators.asp \\ 
  Finland & Regions &  19 &  95 & Statistics Finland (2012). Courtesy of Markus Rapo. \\ 
  France & Departements &  96 & 1042 & INSEE (2006). Donnﾎes de Dﾎmographie rﾎgionale 1954 ﾈ 1999 (CD-ROM). Courtesy of Fabienne Daguet ; INSEE (1998). Les evolutions demographiques departementales et regionales entre 1975 et 2006 ; INSEE (2012). 1990-2009 statistiques de l'ﾎtat civil et estimations de population. Tableay P3D -  indicateurs gﾎnﾎraux de population par dﾎpartement et rﾎgion (sd2010\_p3d\_fe.xls) \\ 
  Germany & States &  16 & 142 & Germany Federal States Working Group (2012). Data provided by Frank Swiaczny at the Bundesinstitut fﾟr Bevﾚlkerungsforschung Redaktion Comparative Population Studies and Sebastian Klﾟsener at MPI Rostock \\ 
  Greece & Regions &  13 &  52 & EuroStat (2012). Fertility rates by age and NUTS 2 regions, demo\_r\_frate2 (http://ec.europa.eu/eurostat/data/database) \\ 
  Hungary & Regions &   7 &  28 & EuroStat (2012). Fertility rates by age and NUTS 2 regions, demo\_r\_frate2 (http://ec.europa.eu/eurostat/data/database) ; Hungarian Central Statistics Office \\ 
  Italy & Regions &  21 & 242 & EuroStat (2012). Fertility rates by age and NUTS 2 regions, demo\_r\_frate2 (http://ec.europa.eu/eurostat/data/database) ; ISTAT (2012). SerieStoriche: Tavola 2.7.1 - Tassi di feconditﾈ totale (TFT) per ordine di nascita, anno e regione - Anni 1952 - 2008 \\ 
  Norway & Counties &  19 & 171 & Statistics Norway (2012). Stat Bank: Table 08556. Total fertility rate and age-specific fertility rates for 5-year periods (http://www.ssb.no/en/table/08556) ; Courtsey of Eva Hoel. \\ 
  Poland & Provinces &  16 &  64 & Central Statistical Office of Poland (2012). Courtesy of Karolina Szlesinger.; EuroStat (2012). Fertility rates by age and NUTS 2 regions, demo\_r\_frate2 (http://ec.europa.eu/eurostat/data/database) \\ 
  Portugal & Regions &   7 &  28 & EuroStat (2012). Fertility rates by age and NUTS 2 regions, demo\_r\_frate2 (http://ec.europa.eu/eurostat/data/database) and Statistics Portugal, INE (2012). Demographic indicators: Total fertility rate (No.) by Place of residence (NUTS - 2002); Annual (http://www.ine.pt) \\ 
  Romania & Counties &  42 & 398 & Romania National Institute of Statistics, Demographic Yearbook 2006 and courtesy of Marcela Postelnicu and Ionica Berevoescu ; Anuarul Statistic al orasului Bucuresti (1866-1930) \\ 
  Russia & Regions &  82 & 525 & Russia in 1946-1958: Andreev E.M., Darsky L.E., Kharkova T.L.  Demographic History of Russia: 1927-1959. Moscow: Informatika, 1998. - 187 p. (in Russian) ; Russia in 1959-2010 and regions in 1989-2010: calculation of Evgeny Andreev based on Russian population statistic data. ; Rosstat for regions for 1978-79, 1982-83, 1984-85, 1986-87, 1988-89. \\ 
  Serbia & Regions &   3 &  36 & Division of demography Statistical Office of the Republic of Serbia (2012). Courtesy of Gordana Bjelobrk. \\ 
  Slovakia & Regions &   8 &  24 & Statistical Office of the Slovak Republic (2012). Courtsey of Jaroslav Sedivy. \\ 
  Slovenia & Regions &   2 &   8 & EuroStat (2012). Fertility rates by age and NUTS 2 regions, demo\_r\_frate2 (http://ec.europa.eu/eurostat/data/database) ; Statistical Office of the Republic of Slovenia (2012). Basic data on live births, statistical regions, Slovenia, annually. (http://www.stat.si) \\ 
  Spain & Auton. Commun. &  17 & 153 & Instituto Nacional de Estadﾒstica (2012). Courtesy of Miguel Angel Martﾒnez Vidal and Fertility rate by Autonomous Community http://www.ine.es/jaxi/menu.do?L=1\&divi=IDB\&his=0\&type=db \\ 
  Sweden & Counties &  21 & 567 & Statistic Sweden (2012). Courtesy of Tomas Johansson and Statistic Sweden (1999). Demographic trends for 250 years: Historical statistics of Sweden \\ 
  Switzerland & Cantons &  26 & 556 & Courtesy of Thomas Spoorenberg ; Wanner P. (2000), "Caractﾎristiques des rﾎgimes dﾎmographiques des cantons suisses 1870-1996", AIDELF, Colloque international de la Rochelle, 22-26 septembre 1998, Paris : Presses Universitaires de France ;  BEVNAT, ESPOP (2012). Encyclopﾎdie statistique de la Suisse. Indicateur conjoncturel de fﾎconditﾎ selon le canton, de 1981 ﾈ 2010 (bfs.admin.ch) \\ 
  Ukraine & Regions &  27 & 133 & State Statistics Service of Ukraine (2012). Courtesy of Mykola Afanasiev. (http://database.ukrcensus.gov.ua/MULT/Database/Population/databasetree\_en.asp and http://www.ukrstat.gov.ua/druk/katalog/nasel/nasel\_2010.zip) \\ 
  United Kingdom & Constit. Countries &   4 &  34 & UK ONS, Vital Statistics Outputs Branch (2012). Courtesy of Matthew Ford. \\ 
   \hline
\hline
\label{tab:datasource}
\end{longtable}

}
 \thispagestyle{empty}
    \end{landscape}
    \clearpage
}

\end{document}